\documentclass[prb,showpacs,twocolumn,aps,superscriptaddress,a4paper]{revtex4-1}
\usepackage{color}
\usepackage{dcolumn}
\usepackage{graphicx}
\usepackage{latexsym}
\usepackage{amssymb,amsmath,bm,bbm}
\usepackage{amsfonts}
\usepackage{color}

\usepackage{array}
\usepackage{graphicx}
\usepackage{ifpdf}
\usepackage{color} 

\DeclareGraphicsExtensions{.pdf,.gif,.jpg,.png}

\newcommand{\ba}{\begin{align}}
\newcommand{\ea}{\end{align}}
\newcommand{\be}{\begin{equation}}
\newcommand{\ee}{\end{equation}}
\newcommand{\bml}{\begin{multline}}
\newcommand{\eml}{\end{multline}}
\newcommand{\bea}{\begin{eqnarray}}
\newcommand{\eea}{\end{eqnarray}}

\def\ppe{P\&p\ }
\def\a{\alpha}
\def\b{\beta}
\def\g{\gamma}
\def\G{\Gamma}
\def\d{\delta}
\def\D{\Delta}
\def\e{\epsilon}

\def\th{\theta}

\def\m{\mu}
\def\n{\nu}

\def\p{\pi}

\def\r{\rho}
\def\rt{\tilde{\rho}}
\def\s{\sigma}
\def\S{\Sigma}
\def\t{\tau}

\def\x{\chi}
\def\w{\omega}
\def\W{\Omega}
\def\q{\psi}
\def\Q{\Psi}

\def\bgh{\mbox{\boldmath $\eta$}}

\def\bld{{\mathbf d}}

\def\ble{{\mathbf e}}

\def\blr{{\mathbf r}}

\def\blE{{\mathbf E}}

\def\callU{\mbox{$\mathcal{U}$}}

\def\iif{\infty}
\def\bra{\langle}
\def\ket{\rangle}

\def\Tr{{\rm Tr}}

\def\Im{{\rm Im}}

\def\1op{\hat{\mathbbm{1}}}
\def\1{\mathbbm{1}}
\def\nn{\nonumber}

\renewcommand{\[}{\left[}
\renewcommand{\]}{\right]}
\renewcommand{\(}{\left(}
\renewcommand{\)}{\right)}

\def\ai		{{\em ab--initio}}

\def\rr		{{\bf r}}

\def\qq		{{\bf q}}

\def\qq		{{\bf q}}

\def\dd         {{\bf d}}

\def\gd         {\delta}

\def\gee        {\epsilon}

\def\go         {\omega}

\def\gr         {\rho}

\renewcommand{\[}{\left[}
\renewcommand{\]}{\right]}
\renewcommand{\(}{\left(}
\renewcommand{\)}{\right)}

\def\dk         {\frac{d\,{\bf k}}{\(2 \pi\)^3}}
\def\dk1        {\frac{d\,{\bf k}_1}{\(2 \pi\)^3}}

\def\dk         {\frac{d\,{\bf k}}{\(2 \pi\)^3}}

\def\bverb      {\renewcommand{\baselinestretch}{1}\begin{verbatim}}
\def\everb  	{\end{verbatim}\renewcommand{\baselinestretch}{1.3}}

\begin{document}

\title{Non--equilibrium Bethe-Salpeter equation for transient photo--absorption spectroscopy}

\author{E. Perfetto}
\affiliation{Dipartimento di Fisica, Universit\`{a} di Roma Tor Vergata,
Via della Ricerca Scientifica 1, 00133 Rome, Italy; and European Theoretical Spectroscopy Facility (ETSF)}
\affiliation{INFN, Laboratori Nazionali di Frascati, Via E. Fermi 40, 00044 Frascati, 
Italy}
\author{D. Sangalli}
\affiliation{Istituto di Struttura della Materia of the National Research 
Council, Via Salaria Km 29.3, I-00016 Montelibretti, Italy; and European Theoretical Spectroscopy Facility (ETSF)}
\author{A. Marini}
\affiliation{Istituto di Struttura della Materia of the National Research 
Council, Via Salaria Km 29.3, I-00016 Montelibretti, Italy; and European Theoretical Spectroscopy Facility (ETSF)}
\author{G. Stefanucci}
\affiliation{Dipartimento di Fisica, Universit\`{a} di Roma Tor Vergata,
Via della Ricerca Scientifica 1, 00133 Rome, Italy; and European Theoretical Spectroscopy Facility (ETSF)}
\affiliation{INFN, Laboratori Nazionali di Frascati, Via E. Fermi 40, 00044 Frascati, 
Italy}

\begin{abstract}
In this work we propose an accurate first-principle approach to calculate the transient 
photo--absorption spectrum measured in Pump\&\,Probe experiments. We 
formulate a condition of {\em adiabaticity} and  thoroughly analyze 
the simplifications brought about by the 
fulfillment of this condition in the non--equilibrium 
Green's function (NEGF) framework. Starting from the Kadanoff-Baym equations 
we derive a non--equilibrium Bethe--Salpeter equation (BSE) for the 
response function that can be 
implemented in most of the already existing {\em ab--initio} codes. 
In addition, the {\em adiabatic} approximation is benchmarked against 
full NEGF simulations in simple model hamiltonians, even under  
extreme, nonadiabatic conditions where it is expected to fail. 
We find that the non--equilibrium BSE is very robust and 
captures 
important spectral features in a wide range of experimental configurations.

\end{abstract}



\maketitle

\section{Introduction}
\label{sec:intro}
The impressive progresses in ultrafast and ultrastrong 
laser-pulse technology has paved the way to the modern non--equilibrium  (NEQ)
attosecond 
spectroscopies.\cite{bgk.2009,spsk.2012,ghlsllk.2013,kc.2014,science.2014} Unlike conventional 
spectroscopies, the sample is 
driven away from equilibrium by a strong laser pulse 
(the pump)  before the photo--absorption of a weaker (perturbative) probe field is measured.
Photo--absorption pump-and-probe\,(P\&p) spectroscopy experiments are carried out using pump 
pulses with frequency in the 
infrared-ultraviolet range and ultrashort probe pulses
(down to a few hundreds of attoseconds).
By varying the delay between the pump and 
probe pulses one can monitor the excited-state dynamics in a wide 
energy range.

For samples of linear dimension
(in the case of extended systems this is the dimension of the primitive cell)
smaller than the wave--length of the incident light,
the measured signal can be calculated theoretically from 
the NEQ density response
function~\cite{bm.1997,wssd,sypl.2011,mpb.2012,blm.2012,dgc.2013,PS.2015} $\x$ (dipole approximation)
or, equivalently, from equilibrium dipole correlators of order larger than 
two.~\cite{Shen-book,kbs.1988,ym.1990,pm.1992,Mukamel-book,yzc.1997}
In the present manuscript we follow the first path.

At equilibrium $\x(\w)$ can be used to construct the dipole--dipole correlation 
function $\alpha\(\w\)$ in isolated system, or the dielectric function $\epsilon\(\go\)$ in extended systems.
For correlated systems the calculation of $\x$ is, in general, a 
difficult task and one has to resort to approximations. 
The most suitable many--body scheme to implement depends on the sample.
For atomic or small molecular samples the Configuration Interaction (CI)
scheme consists in expanding the many--body state in Slater determinants
to obtain eigenstates and eigenvalues.
Subsequently the oscillator strengths are computed and used to
construct $\x$ from a Lehmann representation.
For molecules with tens or more nuclei as well as for crystals the number of 
CI configurations is too large for present-day computational capabilities and 
alternative (statistical in nature) approaches are required. One such 
approach is many--body perturbation theory (MBPT). In MBPT the 
two--particle electron--hole propagator $L$
satisfies a diagrammatic equation known as 
the Bethe--Salpeter Equation (BSE) and $\x$ is constructed from a 
space-time contraction of the arguments of 
$L$.~\cite{strinati_review,svl-book}
The BSE has been successfully applied to study photo--absorption 
spectroscopy of systems ranging from small molecules to bulk metals 
and insulators.
In this context the BSE is solved at the GW level with a statically 
screened interaction.\cite{Strinati,strinati_review,orr.2002,ardo.1998,bsb.1998,rl.1998,pphs.2011}.

Another convenient alternative to CI (and MBPT) is the Linear 
Response (LR)
Time--Dependent Density--Functional Theory (TDDFT).~\cite{pgg.1996,Casida1}
Although TDDFT is in principle exact,~\cite{rg.1984,Ullrich} the available functionals for 
actual calculations are based on the Adiabatic Local Density 
Approximation (ALDA).\cite{rozzi-nature,miguel,degiovannini}
It is well known that ALDA functionals fail in capturing 
double-excitations,\cite{mzcb.2004,kk.2008} charge 
transfer excitations\cite{ngb.2006,m.2005,mt.2006} 
or the Coulomb blockade 
phenomenon\cite{sk.2011,ks.2013} in equilibrium systems.
For extended systems ALDA performs poorly in the description of the
response function as it misses the long--range
electron--hole interaction needed to describe excitons.\cite{orr.2002,roro.2002}
Therefore, the applicability of LR--TDDFT is at present  
restricted to weakly correlated systems with a spectrum dominated by 
single particle-hole excitations. 

Similarly to the equilibrium case the 
\ppe photo--absorption spectrum is described by the NEQ 
response function $\x(t,t')$.
In this work we identify a set of constraints between characteristic times 
that allows us to rewrite  $\x(t,t')$ as a 
function of the delay $\t$ between the pump and
probe pulses and of the time difference $t-t'$, i.e.,
$\x(t,t')=\x^\t(t-t')$.
Henceforth we will refer to this approximation as the {\em 
adiabatic approximation}. 
The mathematical rigorous definition of the adiabatic approximation
as well as its testing in a \ppe set-up is the central objective
of the present manuscript. 

The adiabatic response function can be  computed at different 
levels of accuracy depending on the theoretical scheme used.
In the CI approach the time--dependent expansion coefficients are 
used to calculate the time--dependent 
product of oscillator strengths and subsequently these products are inserted into a 
Lehmann--like representation of the NEQ adiabatic $\x^\t(\w)$
to yield a \ppe spectrum with 
a time--dependent modulation of the 
peak intensity.\cite{rs.2009,nature.2010,sypl.2011,blm.2012}
Within MBPT, instead, we show that the equation of motion for 
$\x^\t(t-t')$ can be rewritten as a BSE. The main difference with the 
equilibrium BSE is that  
the equilibrium single--particle density matrix
is replaced by its time--dependent value as, 
for instance, obtained from the solution of a Boltzmann--like 
equation.~\cite{hj-book} 
The NEQ adiabatic $\x^\t(\w)$ could also be computed within 
LR--TDDFT. However, it is reasonable to expect that the performance 
of ALDA functionals does not improve in NEQ situations.


The structure of the paper is as follows. Section~\ref{sec:tps} 
presents a brief self--contained introduction to the link between the macroscopic 
observable and the microscopic theory. We discuss both the real--time (Section~\ref{sec:tps_real_time}) and the 
response function (Section~\ref{sec:tps_response}) representations. 
Here we also identify a set of characteristic times in terms of which 
the condition of adiabaticity is formulated.
The MBPT approach to $\chi$ is developed in Section~\ref{sec:negf} 
where we introduce the non--equilibrium Green's 
functions (NEGF).\cite{svl-book,hj-book,vLDSAvB.2006,d.1984,bb.2013,kb-book} 
The NEGF approach is computationally more expensive than TDDFT but it has the 
advantage of including dynamical correlations in a nonperturbative 
diagrammatic fashion. To reduce the numerical cost we implement, in Section~\ref{sec:negf_gbka}, NEGF 
within the Generalized Kadanoff--Baym Ansatz~\cite{hj-book,lsv.1986} 
(GKBA) and then derive the linear response equations in 
Section~\ref{sec:negf_linear}. Except for the GKBA no other 
approximations are made at this stage. The complexity of the problem 
is further reduced in Section~\ref{sec:negf_adiabatic}. Here we 
exploit the adiabatic approximation and obtain the central result of 
this work, namely a NEQ--BSE.  We examine differences and 
analogies with the more standard equilibrium BSE and discuss the 
possibility of converting the NEQ--BSE into a Dyson-like equation in 
Section~\ref{sec:negf_dyson}.
Finally, in Section~\ref{sec:numerics} we illustrate the theory in a 
model system by benchmarking the performance of the NEQ--BSE against 
full NEGF calculations.
A summary of the paper and concluding remarks are drawn in Section~\ref{sec:conclusions}.

\section{The transient photo--absorption spectrum}
\label{sec:tps}
In this Section we relate the macroscopic 
quantity measured in a \ppe experiment to 
microscopic quantum-mechanical properties of the probed sample.
This link establishes a connection between the experimental signal 
and the solution
of the complex quantum kinetic equation for the one--particle 
density--matrix. 

\subsection{A real time approach}
\label{sec:tps_real_time}
In a \ppe experiment the transient photo--absorption 
spectrum of a system driven out of equilibrium by a pump field is measured.
The theoretical description of the driven system is achieved by 
evolving the many-body state in the 
simultaneous presence of the pump field and of a weak probe field.
Let $\blE$ and $\ble$ be the electric pump and probe field
respectively. We define the different terms constituting the 
many--body Hamiltonian $\hat{H}(t)$ according to 
\begin{subequations}
\bea
&&\hat{H}^{0}      =\hat{T}+\hat{V}_{N} ,\\
&&\hat{H}^{\rm eq}    =\hat{H}^{0}+\hat{V}_{ee} ,\\
&&\hat{H}^{\rm neq}(t)=\hat{H}^{\rm eq}     + \blE(t)\cdot \hat{\bld}
\label{h+pump},\\
&&\hat{H}(t)          =\hat{H}^{\rm neq}(t) + \ble(t)\cdot \hat{\bld}.
\label{h+probe}
\eea
\label{eq:H_MB_def}
\end{subequations}

\noindent
Here $\hat{T}$ is the kinetic energy operator, ${\hat{V}_N}$
the external static potential of the nuclei and ${\hat{V}_{ee}}$ the 
electron--electron interaction. Therefore $\hat{H}^{\rm eq}$ is the
Hamiltonian of the unperturbed system. The inclusion of other 
interactions, e.g., the electron-phonon interaction, does not modify 
the derivation and the results of the present section. 
The terms ${\blE(t)\cdot \hat{\bld}}$ and ${\ble(t)\cdot \hat{\bld}}$
describe the coupling of the electrons with the pump and probe fields in the dipole 
approximation, $\hat{\bld}$ being the dipole operator (see below for 
its mathematical definition).
For simplicity, we consider linearly polarized pump and probe fields:
\begin{subequations}
\begin{gather}
\blE(t)={\bm \eta}_{P} E(t) ,\\
\ble(t)={\bm \eta}_{p} e(t) ,
\end{gather}
\label{eq:4}
\end{subequations}
with ${\bm \eta}_{P}$ and ${\bm \eta}_{p}$ the polarization vectors. 
The generalization to other kind of polarizations is 
straightforward.

We work in the second quantization formalism and introduce 
a suitable single--particle basis with orthonormal wave--functions $\{\varphi_{i}\(\blr\)\}$. 
Then the creation and 
annihilation field-operator $\hat{\q}^{\dag}\(\blr\)$ and 
$\hat{\q}\(\blr\)$ for a particle at position $\blr$ in space
are expanded according to 
$\hat{\q}\(\blr\)=\sum_{i}\varphi_{i}\(\blr\)\hat{c}_{i}$.
The one--particle density--matrix operator takes the form
\begin{align}
\hat{\gr}(\blr,\blr')=\hat{\q}^{\dag}(\blr)\hat{\q}(\blr')=\sum_{ij}\varphi^*_{i}(\blr) \varphi_{j}(\blr') \hat{\gr}_{ji},
\label{eq:8}
\end{align}
with $\hat{\gr}_{ji}=\hat{c}^{\dagger}_{i}\hat{c}_{j}$. 
Similarly, the dipole operator projected along the probe field in the $\{\varphi_{i}\(\blr\)\}$  
basis reads
\be
\hat{d}={\bm \eta}_{p}\cdot {\hat{\bld}\equiv \int d\blr\; ({\bm \eta}_{p}\cdot \rr)\hat{\gr}(\blr,\blr)}=
d_{ij} \hat{\gr}_{ji},
\label{dipbasis}
\ee
with  $d_{ij}=\int d\blr\,
\varphi^*_{i}(\blr)({\bm \eta}_{p}\cdot \rr)  \varphi_{j}(\blr)$ the dipole matrix elements.
In Eq.~(\ref{dipbasis}) and in the remainder of the paper we use the 
Einstein convention that repeated indices are summed over. 
The time--dependent expectation value of the dipole operator 
is given by
\begin{align}
d\(t\)=\bra\Q(t)|\hat{d}|\Q(t)\ket
=d_{ij}\bra\Q(t)| \hat{\gr}_{ji}|\Q(t)\ket,
\label{dt}
\end{align}
where $|\Q(t)\ket$ is the state of the system at time $t$.

Without any loss of generality we assume that the switch-on time of 
the pump and probe fields is larger than zero; hence the system is in 
the ground state $|\Q_{g}\ket$ at time $t=0$. Let 
$\hat{\callU}_H(t)$  be the unitary evolution operator corresponding 
to a system with dynamics $\hat{H}$
\be
\hat{\callU}_H(t)\equiv Te^{-i\int_{0}^{t}d\bar{t}\; 
[\hat{H}(\bar{t})]},
\quad t>0
\label{eq:evo_def}.
\ee
The time-dependent matrix elements of the one--particle density 
matrix in the presence of both the pump and probe fields are therefore 
given by 
\be
\r_{ji}(t)\equiv \bra\Q(t)| \hat{\gr}_{ji}|\Q(t)\ket=
\bra\Q_{g}|\,\hat{\callU}^{\dag}_H(t)\hat{\gr}_{ji}\,
\hat{\callU}_H(t)|\Q_{g}\ket.
\ee
Replacing $\hat{H}$ with $\hat{H}^{\rm neq}$ in Eq.~\eqref{eq:evo_def}
we have the evolution operator in the presence of the pump
only, ${\hat{\callU}_{H^{\rm neq}}(t)}$.
To simplify the notation we put a tilde on  
time-dependent expectation values obtained with a probe--free 
propagation. Thus 
\be
\tilde{\r}_{ji}(t)
=\bra\Q_{g}|\,\hat{\callU}^{\dag}_{H^{\rm neq}}(t)\hat{\gr}_{ji}\,
\hat{\callU}_{H^{\rm neq}}(t)|\Q_{g}\ket,
\ee
and hence $\tilde{d}(t)=d_{ij}\tilde{\r}_{ji}(t)$.

For optically thin samples~\cite{footnote_thin}
the transmitted probe field $\ble'(t)={\bm \eta}_{p} e'(t)$ 
is related to the probe--induced variation $ \gd\dd\(t\)\equiv 
\bld(t)-\tilde{\bld}(t)$ of the dipole moment by\cite{PS.2015}
\be
\ble'(t)=\ble(t)+\frac{2\p}{S c}\frac{d}{dt} \[\gd\dd\(t\)\],
\label{transe}
\ee
where $S$ is the cross section of the sample (assumed to be 
smaller than the cross section of the laser beam).

The transmitted probe field is typically split in two 
halves and then merged back by a spectrometer, thus generating 
an electric field $\frac{1}{2}[\ble'(t)+\ble'(t-\d)]$ with a tunable 
delay $\d\geq 0$. In a \ppe experiment the integrated intensity of 
this field, i.e., the total absorbed energy per unit area, is 
measured as a function of $\d$:
\be
I'(\d)=\frac{c}{4\p}\int_{-\iif}^{\iif}dt
\left|\frac{\ble'(t)+\ble'(t-\d)}{2}\right|^{2}.
\label{intdelta}
\ee
The resulting function $I'(\d)$ is then cosine-transformed 
\be
\mathfrak{I}'(\w)=\int_{0}^{\iif}  d\d \;I'(\d)\cos(\w\d),
\ee
to gain information about the absorption energies of the system. 
Although the probe pulse has a finite duration the 
time integral in Eq.~(\ref{intdelta}) goes from minus to plus infinity 
since the cosine transform requires $I'(\d)$ for all delays $\d\geq 0$.

Performing an analogous spectral decomposition of 
$\ble(t)$ we get the intensity $\mathfrak{I}(\w)$ of the incident 
probe field. The photo--absorption spectrum $\mathfrak{S}(\w)$ is therefore given by the difference 
\be
\frac{\mathfrak{S}(\w)}{S}=\mathfrak{I}(\w)-\mathfrak{I}'(\w).
\ee
Using Eq.~(\ref{transe}) it is straightforward to show 
that\cite{PS.2015}
\be
\mathfrak{S}(\w)=-2\w\,\Im\left[e\(\w\)\gd d\(\w\)\right]-\frac{2\p}{S c}
\left|\w \gd d(\w)\right|^{2},
\label{eqspec}
\ee
where $e\(\w\)$ and $\d d\(\w\)$ are the Fourier transform of 
the time-dependent probe field $e(t)$ and probe--induced dipole moment 
$\gd d(t)$ respectively. 
This relation expresses the aforementioned link between the macroscopic 
intensity of the transmitted probe field measured in a \ppe experiment
and the microscopic dipole moment.

In photo--absorption experiments of 
equilibrium systems (no pump) the induced electric field [second term 
in the r.h.s. of Eq.~(\ref{transe})] is typically much smaller than 
the incident probe, and the quadratic term in the dipole moment 
appearing in $\mathfrak{S}(\w)$ can be safely discarded. Moreover, 
$\gd d (\w)=\a(\w)e(\w)$ with $\a(\w)\equiv 
d_{ij}\chi_{\substack{ji\\ kl}}(\w)d_{lk}$ having the property that
$\Im[\a(\w)]\lessgtr 0$  for $\w\gtrless 0$
(see next section), and therefore the ratio 
$\mathfrak{S}(\w)/|e(\w)|^{2}$ is {\em positive} 
and {\em independent} of the shape of the probe. On the contrary,
the photo--absorption spectrum of a pump driven system is not 
an intrinsic property of the sample since $\gd d(\w)$, 
although still linear in $e$, depends on 
$e(\w')$ at all possible frequencies $\w'$. Translating 
this statement from frequencies to times, the spectrum depends on the 
shape of the probe and on the NEQ state of the system at 
the time the probe pulse enters the sample (hence on the delay 
between the pump and probe pulses).
Furthermore, 
there might be frequencies for which the spectrum $\mathfrak{S}(\w)$ 
is negative due to a dominance of the stimulated emission over 
absorption.

For a fixed shape of the pump and probe pulse the main interest in 
\ppe experiments is to study the evolution of the spectrum as the 
delay $\t$ between the two pulses is varied. 
Assuming that the quadratic term in $\gd d$, see Eq.~(\ref{eqspec}), is 
small  and taking into account that 
$\gd d(\w)=d_{ij}[\r_{ji}(\w)-\tilde{\r}_{ji}(\w)]$, the 
resulting spectrum reads
\be
\mathfrak{S}^{\t}(\w)=
-2\w\,\Im\left[e^{\ast}(\w)d_{ij}\gd\r_{ji}(\w)\right],
\label{eq:trans_spectrum}
\ee
where we define $\gd\r=\r-\tilde{\r}$.
In Eq.~\eqref{eq:trans_spectrum} we explicitly added to 
$\mathfrak{S}$ a dependence on $\t$ since the probe field as well as 
the time-dependent
density matrix depends on the pump--probe delay. This dependence 
is, in general, rather complex and difficult to interpret. 
As we shall see, 
the calculation of the spectrum as well as its physical 
interpretation are greatly simplified if the adiabatic approximation 
is made.

\subsection{A response--function representation: the adiabatic 
condition}
\label{sec:tps_response}
Equation~\eqref{eq:trans_spectrum} can be rewritten in a different way using linear 
response theory out of equilibrium. Let us introduce the (retarded) 
NEQ response 
function
\begin{multline}
\x_{\substack{ ji\\ lk}}\(t,t'\)=
-i\th\(t-t'\)\\
\times \bra\Q_{g}|
\left[\hat{c}^{\dagger}_{iH}\(t\)\hat{c}_{jH}\(t\),
\hat{c}^{\dagger}_{kH}\(t'\)\hat{c}_{lH}\(t'\)\right]|\Q_{g}\ket
\label{eq:chi_def},
\end{multline}
where $\hat{c}_{iH}(t)\equiv \hat{\callU}^{\dag}_{H^{\rm neq}}(t)\hat{c}_{i}\,
\hat{\callU}_{H^{\rm neq}}(t)$ are fermion operators in the 
Heisenberg picture with respect to the probe--free Hamiltonian $\hat{H}^{\rm neq}$.
To first order in $\ble$ the probe--induced variation of the 
dipole moment reads
\bea
\gd d \(t\)&=&d_{ij}\int dt' \x_{ \substack{ ji \\ lk} }\(t,t'\) 
d_{kl}\,e(t')
\nn\\
&=&
 \int dt' \[ d \circ \x\(t,t'\) \circ d \] e(t').
\label{eq:response_general}
\eea
In Eq.~\eqref{eq:response_general} we introduced a short--hand notation for the
contraction of tensors of different rank. Below we define the four 
types  of contractions, which include the one in Eq.~(\ref{eq:response_general}), that we  
use in the manuscript:
\begin{subequations}
\bea
&&(M\circ V)_{pq}\equiv M_{\substack{ pq\\mn}}V_{nm}, \\
&&(T\circ M \circ V) \equiv T_{pq} M_{\substack{ qp\\ mn}}V_{nm}, \\
&&(M \circ N)_{\substack{ mn\\ rs}}\equiv M_{\substack{ mn\\pq}} 
N_{\substack{ qp\\rs}}, \\
&&[N , V]_{\substack{ mn\\ pq}}=-[V , N]_{\substack{ mn\\ pq}}\equiv N_{\substack{ mi\\pq}} V_{in} -  V_{mi} 
N_{\substack{ in\\pq}}
\label{comm}.
\eea
\label{eq:notation}
\end{subequations}
The rank of the tensors will be clear from the context. 
Notice that Eq.~(\ref{comm}) has the same structure of a commutator since 
the lower indices are fixed. Taking into account 
Eq.~(\ref{eq:trans_spectrum}) we clearly see from
Equation~\eqref{eq:response_general} the relation between the \ppe 
spectrum and the NEQ response function; we can also 
appreciate the complex time dependence introduced by the pump field. 
In fact, in equilibrium (no pump) the response
function reduces to a function of $\(t-t'\)$ due to the invariance 
under time-translations. 
Using this invariance, the linear response relation 
Eq.~(\ref{eq:response_general}) in Fourier space reads 
$\gd d(\w)=\a(\w)e(\w)$ with 
$\a=\(d\circ\chi\circ d\)$, and the ratio 
$\mathfrak{S}(\w)/|e(\w)|^{2}$ becomes independent of the probe. 
As already discussed in the introduction the equilibrium response function 
can be calculated by solving the BSE.

In the time domain the equation for the electron-hole 
propagator $L$ ($\chi$ follows from a space-time contraction of $L$) 
is valid out-of-equilibrium too~\cite{svl-book}
but its numerical solution 
is essentially impossible 
for present-days computational capabilities.
The problem is therefore to find a simple but still
accurate approach to calculate the NEQ $\x$ within MBPT.
For this purpose we will 
extend the {\em equilibrium} BSE to NEQ situations relevant to \ppe  experiments and
provide a sound interpretation of the two--time dependence. 
In the following we refer to this equation as  the NEQ--BSE.

We begin the discussion by introducing two fundamental characteristic 
times that support
the {\em adiabatic approximation}: the key idea is that a NEQ--BSE is 
meaningful whenever the system is substantially frozen
in a NEQ configuration during the measurement process. The  characteristic 
times are
\begin{itemize}
\item[(i)] the time scale $T_P$ of the electron dynamics induced by the pump.
If $\Delta t \ll T_P$, then $\rt\(t+\Delta t\)\approx \rt\(t\)$.
\item[(ii)] the life--time $\t_p$ of the dressed 
probe pulse, which is the duration of the measurement process too.
\end{itemize}
We can formulate the condition of applicability of the 
adiabatic approximation as 
\begin{align}
T_P \gg \t_p.
\label{eq:ad_ansatz}
\end{align}
Equation (\ref{eq:ad_ansatz}) expresses the physical condition
that the probe--free $\tilde{\r}(t)$ has to vary on a time 
scale ($T_{P}$) much longer than
the duration ($\t_p$) of the dressed probe.
Of course for $\t_{p}$ to be smaller than typical electronic time 
scales there should exist decay channels faster than the radiative 
decay. This is the case of solid slabs as well as of thick atomic or 
molecular gases. The following analysis applies to these class of 
systems. 

\begin{figure*}[tbp]
\includegraphics[width=0.9\textwidth]{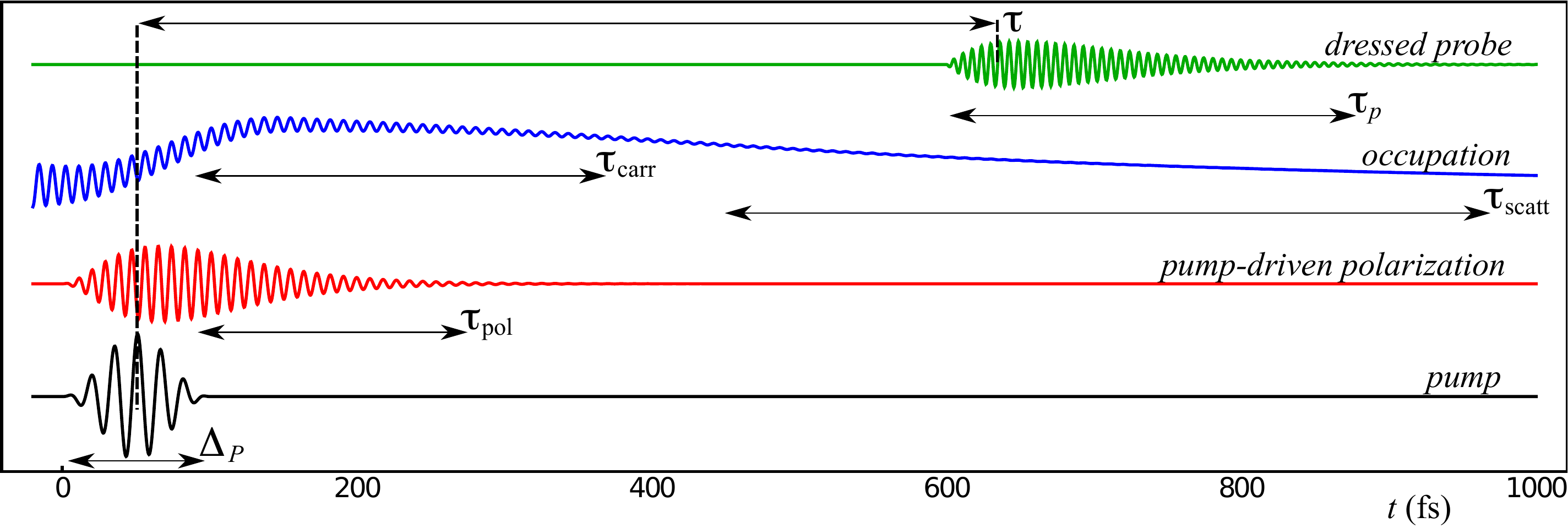}
\caption{Illustration of the characteristic times described in the 
main text: $T_P$ is the time scale of the electron dynamics induced 
by the pump, $\t_{\rm pol}$ the dephasing time of the pump-induced 
polarization, $\t_{\rm carr}$ the stabilization time of the 
occupations, $\t_{\rm scatt}$ the time to relax back to the 
equilibrium state and $\t_{p}$ the life--time of the dressed probe 
field. We also display the delay $\t$ between pump and probe.}
\label{timescales}
\end{figure*}

We identify two different situations where the condition in 
Eq.~(\ref{eq:ad_ansatz}) is fulfilled. 1)
If the pump itself varies on a time scale $T_{P}\gg \t_p$ 
then Eq.~(\ref{eq:ad_ansatz}) is always fulfilled since 
the pump-induced dynamics cannot be faster than $T_{P}$. In this 
case the adiabatic 
approximation, and hence the NEQ--BSE, can be used to 
describe the transient spectrum for any delay $\t$ between the pump and 
probe fields. 2) In general, however, the pump is a pulse of duration 
$\D_{P}$, see Fig.~\ref{timescales}, no longer than a few hundreds of femtoseconds capable of 
inducing arbitrary fast processes. During the action of the pump 
the level occupations change and the system polarizes. Shortly after 
$\D_{P}$ we have a transient period characterized by a 
dephasing-driven drop of the 
pump-induced polarization, we 
denote by $\t_{\rm pol}$ the 
polarization life--time in this nonequilibrium situation, and 
by a stabilization of the level occupations at some nonequilibrium 
value, we denote by $\t_{\rm carr}$ the 
characteristic time for the occupations to stabilize, see again Fig.~\ref{timescales}. Thus, after a 
time $\t_{\rm max}={\rm max}(\t_{\rm pol},\t_{\rm carr})$, typically 
$\t_{\rm pol}<\t_{\rm carr}$, we may say that the system is in a 
quasi--stationary state with carriers in some excited levels. 
In this quasi--stationary regime  
the time to relax back to the ground state is dictated by 
scattering processes (electron--electron, electron--phonon and electron--photon)
and can be of the order of picoseconds. 
If we denote by $\t_{\rm scatt}$ this relaxation time-scale then  
we have $T_{P}=\t_{\rm scatt}$. Suppose now to probe the system 
in this quasi--stationary state  
with a pulse $e(t)$ of duration $\D_{p}$. The probe induces a 
polarization $\d d(t)$ which dresses the bare $e(t)$ and, in general, 
has a finite life--time 
$\tilde{\t}_{p}$. Hence the duration of the dressed probe field, 
which coincides with the duration of the measurement process, is 
$\t_{p}=\D_{p}+\tilde{\t}_{p}$. In this regime the condition in 
Eq.~(\ref{eq:ad_ansatz}) is fulfilled provided that  $\t_{p}$ is shorter than the 
relaxation time $\t_{\rm scatt}$. This is often the case as $\t_{p}$ 
is typically in the femtosecond range.

In Fig.~\ref{timescales} we represent the dressed probe field 
with oscillations of frequency $\w_{p}$.
Although the characteristic frequency $\w_{p}$ can be any, it is 
clear that it is only for 
\begin{align}
\t_p \gg \frac{2\pi}{\w_p}
\label{eq:ft_ansatz}
\end{align}
that the Fourier transform of the probe--induced dipole
has a well defined structure in 
$\w_{p}$. This implies that the life--time $\t_{p}$ also sets a lower 
limit to the frequency resolution of a transient 
spectroscopy experiment. 

When the inequality of Eq.~~(\ref{eq:ad_ansatz}) is satisfied 
the probe sees a NEQ frozen 
system. If we take $t=0$ as the time at which the pump is on then 
the probe acts at $t=\t$ and for times  $(t,t') \in [\t-\t_{p} , \t+\t_{p}]$ the response function
\be
\x(t,t')\approx\x^\t(t-t')
\label{eq:equil_limit}
\ee
depends only on $(t-t')$ to a large extent. 
We will provide a more precise definition of $\x^\t(t-t')$ in 
the next section. For the time being we observe that 
whenever we can make the adiabatic approximation of
Eq.~(\ref{eq:equil_limit}) the transient photoabsorption spectrum 
of Eq.~(\ref{eq:trans_spectrum}) can be written as
\be
\mathfrak{S}^{\t}(\w)=
-2\w\,|e(\w)|^{2}\,\Im\left[d\circ \chi^{\t}(\w)\circ d 
 \right] .
\label{transspectrum}
\ee
Consequently the ratio 
$\mathfrak{S}^{\t}(\w)/|e(\w)|^{2}$ becomes independent of 
the probe and can be interpreted as an {\em intrinsic} property of 
the {\em nonequilibrium} 
system.

As a very general remark we notice that when the system is probed 
after the pump (no overlap between the pulses)
the probe--induced dipole moment 
oscillates at frequencies $\W_{\a\b}=E_{\a}-E_{\b}$, where 
$E_{\a},E_{\b}$ are eigen--energies of $\hat{H}^{\rm eq}$, 
see Eqs.~(\ref{eq:chi_def}) and 
(\ref{eq:response_general}).\cite{PS.2015,FLSM.2015} 
Furthermore, the amplitude of the oscillations depends on the delay 
$\t$.\cite{PS.2015} Therefore
P\&p spectra are richer than equilibrium spectra where
the probe--induced dipole moment 
can oscillate only at frequencies $\W_{\b}=E_{g}-E_{\b}$, with 
$E_{g}$ the ground state energy, with constant amplitudes. The extra transitions are usually referred
to as photo--induced absorption and stimulated emission.

To summarize, Eqs. (\ref{eq:trans_spectrum})  and (\ref{transspectrum})  
represent two different ways of calculating the 
transient photo--absorption spectrum. We could either 
perform a time propagation with both the pump and the probe, a second 
time propagation with only the pump and then extract the probe--induced 
dipole moment $\gd d$ or
we can evaluate the response function 
$\x^{\t}$ from a NEQ--BSE. The latter approach is developed in the 
next section.

\section{A non--equilibrium Green's function approach to transient absorption}
\label{sec:negf}
In the previous section we have introduced the theoretical description of transient absorption
experiments with two possible approaches. The first, which is exact, based on Eq.~\eqref{eq:trans_spectrum}
and the second, which uses the adiabatic approximation, based on Eq.~\eqref{transspectrum}. 
However, these equations assume that it is ideally possible to
compute the exact time-dependent density
matrix or the exact adiabatic response function. This is not doable 
in practice and one has to resort to approximations.
In the following we show how to use NEGF theory to 
obtain a MBPT equation for  $\delta d\(t\)$. In the next Section we 
use this result to generate an equation for the 
NEQ response function $\x(t,t')$, and subsequently 
make the adiabatic approximation to 
derive the NEQ--BSE for $\chi^{\t}(t-t')$.

In the MBPT approach the
description in terms of the many--body hamiltonian containing the electron--electron
interaction is replaced by a description in terms of the one particle 
hamiltonian and the
many--body self--energy. We thus define:
\begin{subequations}
\begin{eqnarray} 
&&h^0            =\left[-\frac{\nabla^{2}}{2} +V_{N}(\blr)\right], \\
&&h^{\rm eq}     =h^0 + \S^{0},                            \\
&&\tilde{h}^t(t) =h^{\rm eq} + \Delta\tilde{\S}_{s}^{t} +\blE(t)\cdot\bld,           \\
&&h^t(t)         =\tilde{h}^{t}(t) + \d\S_{s}^{t} +\ble(t)\cdot\bld.
\end{eqnarray}
\label{eq:h_mbpt_def}
\end{subequations}
We have here introduced the $t$ superscript
to indicate a quantity whose time--dependence is given 
by the implicit dependence on the density matrix.
This means that, for a generic function $f$, we have
\begin{gather}
f^t\equiv f\[\gr\(t\)\],\\ 
f^t(t)\equiv f\[\gr\(t\)\](t).
\end{gather} 
The difference between 
$f^t$ and $f^t\(t\)$ is  that the second function has an explicit 
time dependence too.
Furthermore, to indicate that the function is calculated at the 
probe--free density matrix $\tilde{\r}$ we put a tilde symbol on the 
function. Thus $\tilde{f}^{t}\equiv f\[\tilde{\gr}\(t\)\]$ and 
$\tilde{f}^t(t)\equiv f\[\tilde{\gr}\(t\)\](t)$.
Let us define the three 
different self--energies appearing in Eqs.~(\ref{eq:h_mbpt_def}).
The self--energy $\S^{0}=\S_{s}[\r^{\rm eq}]$ is the static part of the 
equilibrium many--body self--energy and it is therefore calculated at 
the equilibrium density matrix $\r^{\rm eq}$. The self--energies
 ${\D\tilde{\S}_{s}^{t}}$ and ${\d\S_{s}^{t}}$ are the variations due
to a change in $\r$ induced by the pump and the probe respectively:
\begin{subequations}
\begin{gather} 
\D\tilde{\S}_{s}^{t}\equiv \S_{s}[\rt(t)]-\S_{s}[\r^{\rm eq}]=\tilde{\S}_{s}^{t}-\S^{0}, \\
\d\S_{s}^{t}\equiv\S_{s}[\r(t)]-\S_{s}[\rt(t)]=\S_{s}^{t}-\tilde{\S}_{s}^{t}.
\end{gather} 
\label{eq:static_DS_def}
\end{subequations}

In general 
$\S_{s}$ is the Hartree--Fock (HF) plus static correlation self--energy.
It plays a crucial role as it renormalizes the single--particle level energies and
introduces correlation effects (like electron--hole attraction) 
also in the polarization function. The different possible 
approximations to $\S_{s}$
reflect the different kind of physics introduced in the dynamics:
\begin{itemize}
\item[(i)] A mean--field potential that mimics the correlation effects. An example is
DFT where $\S_{s}^{t}$ is local in space and given by the sum of the 
Hartree and exchange--correlation potential.
\item[(ii)] HF self--energy. In this case no correlation 
is included. The HF self--energy reads ${\S^{t}_{s}=V \circ \r(t)}$,
with the four-index tensor ${V_{\substack{ ij\\ mn}}=2v_{imnj}-v_{imjn}}$ and 
$v_{imnj}$ the two-electron Coulomb  integrals.
\item[(iii)] Hartree plus a Coulomb Hole and Screened Exchange\,(COHSEX) self--energy.
In this case correlation is included using a linear--response 
approximation but dynamical effects are neglected. The COHSEX 
self--energy reads
$\S_{s}^{t}=V^{t} \circ \r +W^{t}_C$
with $V^{t}_{\substack{ ij\\ mn}}=2v_{imnj}-v^{t}_{imjn}$ and $W^{t}_C$ the Coulomb hole
potential. In $V^t$  the screened exchange interaction reads 
\begin{align}
v^t(\blr,\blr') \equiv \int\,d\overline{\rr}\, \e^{-1}_{\rm RPA}[\r(t)]\(\rr,\overline{\rr}\) v\(\overline{\rr}-\rr'\).
\end{align}
\end{itemize}

In all cases the static self--energy is a {\em time-local} functional of the 
density matrix.

\subsection{Real--time dynamics (I): the Generalized Kadanoff-Baym Ansatz}
\label{sec:negf_gbka}
In NEGF theory the key quantities are the lesser, ${G^<(t,t')}$, and 
greater, ${G^>(t,t')}$,
Green's functions. 
These functions are defined according to
\begin{subequations}
\begin{gather}
G_{ij}^{<}(t,t')= i\langle \hat{c}^\dag_{jH}(t')\hat{c}_{iH}(t) \rangle    \label{g<_def},  \\
G_{ij}^{>}(t,t')=-i\langle \hat{c}_{iH}(t)\hat{c}^\dag_{jH}(t') 
\rangle    \label{g>_def}.
\end{gather}
\label{g<>_def}
\end{subequations}
It is easy to verify that the one-particle density matrix is given by the lesser Green's 
function at equal times, 
${\gr\(t\)\equiv -i G^{<}\(t,t\)}$. The 
functions $G^{\lessgtr}$ satisfy a set of coupled equations known as the Kadanoff-Baym 
equations 
(KBE).\cite{kb-book,hj-book,vLDSAvB.2006,d.1984,svl-book,bb.2013,DvL.2006,MSSvL.2009} 
The KBE are integro-differential equations with a 
self--energy kernel depending on both $G^{<}$ and $G^{>}$. It is 
possible to collapse the KBE into a single equation for the 
one--particle density 
matrix by making the so called Generalized Kadanoff-Baym Ansatz 
(GKBA).\cite{lsv.1986} The corresponding equation for $\rho$ reads
\be
\frac{d}{dt}\r(t)+i\left[h^t(t),\r(t)\right]=-I^{t}(t),
\label{eq:kbe_general}
\ee
where $h^t(t)$ is defined in Eq.~\eqref{eq:h_mbpt_def}.

The collision integral $I^{t}(t)=I[\r(t)](t)$ on the r.h.s. of Eq. (\ref{eq:kbe_general}), 
at difference with the static self--energies previously discussed, is
non--local in time (unless specific approximations are made). 
The functional form is uniquely determined through the GKBA once an approximation for the 
correlation self--energy, $\S_{c}$, is made. Let us show 
how to obtain $I^{t}$ starting from its exact KBE expression and then 
making the GKBA. From the KBE we have
\begin{multline}
I(t)=\int d\bar{t}\left[\S_{c}^{<}(t,\bar{t})G^{\rm (a)}(\bar{t},t) \right. \\ \left.
+\S_{c}^{\rm (r)}(t,\bar{t})G^{<}(\bar{t},t)\right]+{\rm H.c.}
\label{I}
\end{multline}
with $\S_{c}$ a functional of $G^{<}$ and $G^{>}$. 
The functional form of $\S_{c}$ must be consistent with the choice of $\S_{s}$, i.e.
${\S_{c}=\S-\S_{s}}$ with $\S$ the full many--body self--energy.
Retarded/advanced functions carry a superscript (r)/(a) and are 
defined in terms of the lesser and greater functions according to
\begin{eqnarray}
X^{\rm (r)}(t,t')&=&[X^{\rm (a)}(t',t)]^{\dag} \nonumber  \\
                 &=&\th(t-t')\left[X^{>}(t,t')-X^{<}(t,t')\right],
\end{eqnarray}
where $X$ can be $G$, $\S_{c}$ or any other two-time correlator. The 
GKBA is an ansatz for $G^{\lessgtr}$ which turns $\S_{c}$, and hence 
the collision integral, into a 
functional of $\r$ and $G^{\rm (r)/(a)}$:
\begin{subequations}
\bea
G^{<}(t,t')=&-&G^{\rm (r)}(t,t')\r(t')+\r(t)G^{\rm (a)}(t,t')             \label{g<}, \\
G^{>}(t,t')=&+&G^{\rm (r)}(t,t')\bar{\r}(t')-\bar{\r}(t)G^{\rm (a)}(t,t')  \label{g>},
\eea
\label{g<>}
\end{subequations}
where $\bar{\r}=1-\r$. To transform $I(t)$ into a functional 
$I^{t}(t)$ of the density matrix, and hence to close Eq. 
(\ref{eq:kbe_general}), one needs 
to express  the propagator $G^{(\rm r)}$ in terms of $\r$. Depending on the system there exist optimal 
approximations to the propagator, the most common one being the quasi--particle (QP)
propagator
\be
G^{\rm (r)}(t,t')=-i\th(t-t')Te^{-i\int_{t'}^{t}d\bar{t}\,h^{\rm qp}(\bar{t})}.
\label{gret}
\ee
For (small) finite systems
the choice ${h^{\rm qp}=h^{\rm eq}}$ (usually $h^{\rm eq}$ is the 
HF single--particle hamiltonian)
is a good choice. For extended systems, 
however, the lack of damping in
$h^{\rm eq}$ prevents the system to relax.
In these cases the propagator is typically corrected by adding  
non--hermitean terms given by the quasi--particle 
life--times ${h^{\rm qp}=h^{\rm 
eq}+i\g}$.\cite{marini.2008,marini.2003,h.1992,bsh.1999,m.2012,LPUvLS.2014}

\subsection{Real--time dynamics (II): the linear regime}
\label{sec:negf_linear}
\label{bsesec}
If the probe is a weak perturbation we can work within a linear--response approach. 
Then $\d \S_{s}^{t}$ is of first order in $\ble$ and the 
collision integral can be expanded as
\be
I^{t}(t) \approx \tilde{I}^{t}(t)+\d I^{t}(t).
\label{eq:S_I_linearization}
\ee
Inserting Eq.~\eqref{eq:S_I_linearization} into 
Eq.~\eqref{eq:kbe_general} and equating terms of the same order in 
the probe field we get two equations, one for $\tilde{\r}$ and another 
for $\d\r$ (omitting the explicit time-dependence from the various 
quantities): 
\begin{subequations}
\begin{gather}
\frac{d}{dt} \rt+   i[\,\tilde{h}^{t},\rt\,]=-\tilde{I}^{t},                               
\label{eq:kbe_lin_r}    \\
\frac{d}{dt} \d\r+  i[\,\tilde{h}^{t},\d\r\,]
+[\,\d\S^{t}_{s}+\ble\cdot\bld,\rt\,]=-\d I^{t}.
\label{eq:kbe_lin_delta_r}
\end{gather}
\label{eq:kbe_lin}
\end{subequations}
As we are in the linear--response regime, 
we can rewrite $\d\S^t_{s}$ and $\d I^{t}$ in terms of kernel 
functions of the probe--free
density matrix $\tilde{\r}$. The notation introduced proves now useful because
it highlights the dependence on $\tilde{\r}$ and $\d\r$:
\begin{subequations}
\bea
&&\d \S^{t}_{s} = \tilde{K}^{t}_{s} \circ \d\r(t), \\
&&\d I^{t}(t) = \int d\bar{t}\, \tilde{K}_{c}^{t}(t,\bar{t}) \circ 
\d\r(\bar{t}) .
\eea
\label{eq:S_I_lin}
\end{subequations}
The static kernel $\tilde{K}^{t}_{s}$ and the correlation kernel $\tilde{K}_{c}^t$ depend only on 
$\tilde{\r}$. Furthermore $\tilde{K}_{c}^t(t,\bar{t})$ vanishes for $\bar{t}>t$ 
since $I^{t}(t)$
depends  on $\r(\bar{t})$ only for $\bar{t}<t$, as it follows directly from Eq.~\eqref{I}
and the GKBA in Eqs.~\eqref{g<>}.
With Eqs.~\eqref{eq:S_I_lin} we can rewrite
Eq.~\eqref{eq:kbe_lin_delta_r} as
\begin{multline}
\frac{d}{dt}\d\r
+i[\, \tilde{h}^t ,\d\r\, ]  
+i[\, \tilde{K}^{t}_{s}\circ\d\r+\ble\cdot\bld ,\rt\,] = \\
 -\int d\bar{t}\, \tilde{K}_{c}^{t}(t,\bar{t}) \circ \d\r(\bar{t}).
\label{eq:kbe_lin_delta_r_k}
\end{multline}
Equation~\eqref{eq:kbe_lin_delta_r_k} is the many--body equation for the 
calculation of the probe--induced change of the density matrix.
In the next section we combine Eq.~\eqref{eq:kbe_lin_delta_r_k} with the condition of adiabaticity in 
Eq.~(\ref{eq:ad_ansatz}) to derive  a NEQ--BSE.

\section{Non--equilibrium Bethe--Salpeter equation}
\label{sec:negf_adiabatic}
The next step in the derivation of a BSE in the presence of the pump field is 
to transform Eq.~\eqref{eq:kbe_lin_delta_r_k} into an equation for the response function.
To this end we use the relation:
\be
\chi_{\substack{ ji\\ lk}}(t,t')=\frac{\d\r_{ji }(t)}{\d u_{kl}(t')},
\ee
with ${u_{kl}(t)=\ble(t)\cdot \bld_{kl}}$. 
Taking the functional derivative of 
Eq.~\eqref{eq:kbe_lin_delta_r_k}  with respect to $u(t')$ and find
\begin{multline}
\frac{d}{dt}\x(t,t')+i\left[\tilde{h}^t(t),\x(t,t')\right]+\\
+i\left[\tilde{K}^{t}_{s}\circ\chi(t,t')+\1 \d(t-t'),\rt(t)\right]=\\
  -\int d\bar{t}\; \tilde{K}^{t}_{c}(t,\bar{t})\circ \x(\bar{t},t'),
\label{eq:kbe_lin_chi}
\end{multline}
where we introduced the four index tensor $\1_{\substack{ ji\\ 
kl}}=\d_{jl}\d_{ik}$. At zero pump this equation reduces to the 
equilibrium BSE
\begin{multline}
\frac{d}{dt}\x^{\rm eq}(t-t')+i\left[h^{\rm eq},\x^{\rm eq}(t-t')\right]+\\
+i\left[K^{\rm eq}_{s}\circ\chi^{\rm eq}(t-t')+\1 \d(t-t'),\r^{\rm eq}\right]=\\
  -\int d\bar{t}\; K^{\rm eq}_{c}(t-\bar{t})\circ \x^{\rm eq}(\bar{t}-t'),
\label{eq:kbe_lin_chi_eq}
\end{multline}
The differences between Eq.~\eqref{eq:kbe_lin_chi} and Eq.~\eqref{eq:kbe_lin_chi_eq} are:
\begin{itemize}
\item[(i)] in the equilibrium limit all quantities depend on the 
relative time coordinate only;
\item[(ii)] $\r^{\rm eq}$ is time independent while $\rt(t)$ is time dependent;
\item[(iii)] the static equilibrium hamiltonian ${h^{\rm eq}}$ is replaced by the time dependent ${\tilde{h}^t(t)}$;
\item[(iv)] the kernels ${\tilde{K}^{t}_{s}}$ and 
${\tilde{K}_{c}^{t}(t,\bar{t})}$ are evaluated at the pump--driven 
time-dependent density matrix ${\rt}$ whereas the kernels 
${K^{\rm eq}_{s}}$ and 
${K^{\rm eq}_{c}(t-\bar{t})}$ are evaluated at the static equilibrium 
density matrix $\r^{\rm eq}$. 
\end{itemize}
Due to these points it is not possible to reduce Eq.~\eqref{eq:kbe_lin_chi} to
an algebraic equation for ${\x(t,t')}$, as it is commonly done in 
state-of-the-art  equilibrium calculations 
after Fourier transforming with respect to the time difference ${t-t'}$.
In Eq.~\eqref{eq:kbe_lin_chi} ${\x(t,t')}$ is not 
a function of ${t-t'}$ and, furthermore, the dependence on $t$ appears 
both implicitly and explicitly in $\tilde{h}^t$, 
$\tilde{K}^{t}_{s}$, $\tilde{K}^{t}_{c}$ and $\tilde{\r}$.

Analytical progress can be made provided that the adiabatic 
condition, see
Eq.~(\ref{eq:ad_ansatz}), is fulfilled. We recall that in this 
approximation the pump--driven density matrix $\rt(t)$ varies slowly 
over the 
life--time $\t_{p}$ of the dressed probe. Thus for $t\in [\t-\t_{p} , \t+\t_{p}]$
we have
\be
\rt(t)\approx \rt(\t).
\label{eq:aa_app_1}
\ee
In the same time window $\blE(t)\approx \blE(\t)$  and hence  Eq.~(\ref{eq:aa_app_1})  
implies that
\begin{gather}
\tilde{h}^t(t)\approx\tilde{h}^\t(\t),\\
\tilde{K}^{t}_{s}\approx \tilde{K}^{\t}_{s}.
\end{gather}
This is a direct consequence of the fact that
the functionals $\tilde{h}^t(t)$ and $\tilde{K}^{t}_{s}$ are time-local functionals of ${\rt(t)}$. 

Another simplification brought about by the adiabatic condition is that for 
times $t,t'\in  [\t-\t_{p} , \t+\t_{p}]$ the retarded Green's 
function, see Eq.~(\ref{gret}), can be approximated as
\be
G^{\rm (r)}(t,t')\approx -i\th(t-t')\exp[-ih^{\rm qp}(\t)(t-t')] .
\label{eq:aa_app_3}
\ee
Therefore,  the adiabatic retarded Green's function is 
 invariant under time 
translations. The crucial consequence of this fact is that
the correlation kernel too becomes a function of the time difference 
only:
\be
\tilde{K}^{t}_{c}(t,t')\simeq \tilde{K}^{\t}_{c}(t-t').
\label{eq:aa_app_4}
\ee
Taking into account Eqs.~(\ref{eq:aa_app_1}--\ref{eq:aa_app_4})
we see that the solution of Eq.~\eqref{eq:kbe_lin_chi} is 
a response function ${\x(t,t')\simeq \x^{\t}(t-t')}$ depending 
on the delay $\t$ and on the time 
difference $t-t'$.

In the adiabatic approximation Eq. (\ref{eq:kbe_lin_chi}) can be 
conveniently Fourier transformed to yield
 an algebraic 
equation for the frequency-dependent response function
\begin{multline}
-i\w\x^{\t}(\w)+i\left[\tilde{h}^\t(\t),\x^{\t}(\w)\right]
+i\left[\tilde{K}^{\t}_{s}\circ\chi^{\t}(\w)+\1 ,\rt(\t)\right]\\=
  - \tilde{K}^{\t}_{c}(\w)\circ \x^{\t}(\w).
\label{eq:kbe_lin_chi_adiab}
\end{multline}
This is the aforementioned NEQ--BSE and the main result 
of the present work. We emphasize that $\chi^{\t}$ is the response 
function of the finite system.
In the case of extended systems $\chi^{\t}$ is equivalent to the 
macroscopic response function obtained from a supercell calculation where the spatial long--range
component of the induced Hartree field (corresponding to its $\qq\rightarrow 0$ Fourier component) has been 
removed~\cite{orr.2002}.

The solution of Eq. (\ref{eq:kbe_lin_chi_adiab}) requires a preliminary calculation 
of the one-particle density matrix $\tilde{\r}(t)$.
In the next sub--section we show how to rewrite the NEQ--BSE as a Dyson equation
for $\chi^{\t}$. The NEQ Dyson equation is then compared with its 
equilibrium counterpart
to provide an intuitive physical interpretation of the response 
function.

\subsection{Reduction to a Dyson equation}
\label{sec:negf_dyson}
The NEQ--BSE, Eq.~\eqref{eq:kbe_lin_chi_adiab}, can be implemented in most
of the \ai\, numerical schemes and codes. However, in order to create 
an even closer connection to 
standard implementations of the BSE we further discuss the approximations and conditions under which
Eq.~\eqref{eq:kbe_lin_chi_adiab} turns into a simple Dyson equation.

The crucial aspect is the choice of the reference basis and its link 
with the adiabatic approximation. Let us first re-examine the 
equilibrium case.  Consider the representation
in which ${h^{\rm eq}}$ is diagonal, i.e., ${h^{{\rm 
eq}}_{ij}=\d_{ij}\e_{i}^{\rm eq}}$.
Then, the equilibrium density matrix is diagonal too
and its entries are the occupation factors of the electronic levels: 
$\r^{\rm eq}_{ij}=\d_{ij}f^{\rm eq}_i$. In this basis 
the Fourier transform of the equilibrium BSE, i.e. 
Eq.~(\ref{eq:kbe_lin_chi_eq}), reads
\begin{multline}
\left[\w\1 - \D\e^{\rm eq} + i K^{\rm eq}_{c}(\w) \right]\circ  \x^{\rm eq}(\w) =  \\
- \D f^{\rm eq} \circ \left[\1+K^{\rm eq}_{s}  \circ\x^{\rm eq}(\w)\right],
\label{eq:kbe_lin_chi_eq_w}
\end{multline}
where 
\be
(\D\e^{\rm eq})_{\substack{ij\\pq}}=(\e_{i}^{\rm eq}-\e_{j}^{\rm 
eq})\1_{\substack{ij\\pq}},
\label{detensor}
\ee
and
\be
(\D f^{\rm eq})_{\substack{ij\\pq}}=(f_{i}^{\rm eq}-f_{j}^{\rm 
eq})\1_{\substack{ij\\pq}}.
\label{dftensor}
\ee
Introducing the response function
\be
\x_0^{\rm eq}(\w)\equiv -\left[\w\1 - \D\e^{\rm eq} + i K^{\rm eq}_{c}(\w) 
\right]^{-1}\circ \D f^{\rm eq}  
\label{chi0eq}
\ee
we can rewrite Eq.~(\ref{eq:kbe_lin_chi_eq_w}) in the form normally used in first principles calculations
\be
\x^{\rm eq}(\w)=\x_0^{\rm eq}(\w)+\x_0^{\rm eq}(\w) \circ K^{\rm eq}_{s} \circ 
\x^{\rm eq}(\w) \ .
\label{chieq}
\ee
The correlation kernel $K^{\rm eq}_{c}(\w)$ appearing in $\x_0^{\rm 
eq}$ deserves a comment. In most of the applications 
$K^{\rm eq}_{c}$ is usually replaced by  a constant, i.e.
$K^{\rm eq}_{c}(\w)\approx \eta$. More sophisticated approximations 
with $(K^{\rm eq}_{c})_{\substack{ij\\mn}}\approx 
\g_i\d_{ij}+\g_m\d_{mn}$, have been explored.~\cite{marini.2008} 
In this case the quasi--particle line--widths $ \g_i$ are 
calculated from equilibrium MBPT.
The approximation of a static correlation kernel is based
on the observation that dynamical corrections to the screened interaction are partially cancelled by the 
dynamical effects in the quasi--particle corrections, see Ref.~\onlinecite{marini.2003}. 

Let us now consider the NEQ--BSE, i.e. Eq.~\eqref{eq:kbe_lin_chi_adiab}.
Like in the equilibrium case we would like to introduce a $\chi_{0}$ 
and turn Eq.~\eqref{eq:kbe_lin_chi_adiab} into a Dyson equation.
However, in the NEQ case neither ${\rt(\t)}$ nor 
${\tilde{h}^{\t}(\t)}$ are diagonal in the eigenbasis of $h^{\rm 
eq}$. Of course we can rotate the equilibrium basis so to have 
${\tilde{h}^{\t}(\t)}$ diagonal
but, in general, ${\rt(\t)}$ has off-diagonal entries in this new 
basis too. 
Let  $O(\t)$ be the orthogonal matrix of the transformation from the equilibrium basis to the 
{\em adiabatic basis} in which  
${\tilde{h}^{\t}(\t)}$ is diagonal
\be
\[O^{\dagger}(\t) \tilde{h}^\t O(\t)\]_{ij}=\gd_{ij}\tilde{\gee}_i\(\t\).
\ee
The NEQ--BSE Eq.~(\ref{eq:kbe_lin_chi_adiab}) in the 
adiabatic basis reads
\begin{multline}
\left[\w\1 -\D \tilde{\e}(\t) +i \tilde{K}^{\t}_{c}(\w)\right] \circ \x^\t(\w) = \\
-\left[\rt(\t),\tilde{K}^{\t}_{s}\circ\chi^{\t}(\w)+\1 \right]
\end{multline}
where the four-index tensor $\D\tilde{\e}(\t)$ is defined as in 
Eq.~(\ref{detensor}) with $\e_{i}^{\rm eq}\to 
\tilde{\e}_{i}(\t)$. Next we define the NEQ 
response function $\chi_{0}^{\t}$ according to
\be 
\x_0^\t(\w)\equiv -\left[\w-\D\tilde{\e}(\t)+i\tilde{K}^{\t}_{c}(\w)\right]^{-1} 
\circ \left[\,\rt(\t),\1 \right]
\label{eq:neq_Xo}
\ee
which generalizes Eq.~(\ref{chi0eq}) to nondiagonal density matrices. 
Using the identity
\be
\left[\rt(\t),\tilde{K}^{\t}_{s}\circ\chi^{\t}(\w)\right]=
\left[\,\rt(\t),\1 \right]\circ\tilde{K}^{\t}_{s}\circ\chi^{\t}(\w)
\ee
we can rewrite the NEQ--BSE in a Dyson-like form
\be
\x^\t(\w)=\x_0^\t(\w)+ \x_0^\t(\w) \circ 
\tilde{K}^{\t}_{s} \circ \x^\t(\w).
\label{eq:neq_bse}
\ee

The analogy between Eq.~(\ref{eq:neq_bse}) and the standard, equilibrium BSE becomes more evident 
if we make some further approximations that are
often used in actual implementations:
\begin{itemize}
\item[(i)] ${h^{\rm eq}}$ constructed from the dynamical GW 
self--energy. In this way the equilibrium basis is the quasi--particles 
basis whose states are renormalized by 
dynamical effects. The same approximation applies, for internal consistency,
to $\D\S^{\t}_{s}$.
\item[(ii)]  In order to recover the equilibrium limit of standard 
BSE implementations
the term ${\d\Sigma^{\t}_{s}}$
must be the statically screened COHSEX approximation. 
This has been proved in Ref.~\onlinecite{Att.2011}.
\end{itemize}
With these two approximations in mind we discuss
Eq.~\eqref{eq:kbe_lin_chi_adiab} in the case of a weak pump field. 
This condition is often 
realized in \ppe experiments as it allows to photo--excite the 
system  without changing too much its electronic and optical 
properties. Therefore, weak pump fields provide a   
non--invasive method to monitor the excited states of the equilibrium system.

For weak pump fields the density of the excited carriers is small. 
This implies that we can approximate the orthogonal matrix
$O_{ij}(\t)\approx \d_{ij}$. In other words the 
adiabatic basis and the equilibrium  basis are essentially the same.
The obvious and physically intuitive consequence of this fact is that 
the diagonal elements of the density matrix are
the NEQ occupations ${\rt_{ii}(\t)=\tilde{f}_i(\t)}$ whereas the 
off-diagonal elements
describe the polarization of the system. 
If the photo--excited carrier density is small the off--diagonal 
elements can be neglected and Eq.~(\ref{eq:neq_Xo}) simplifies to
\be 
\tilde{\x}_0^\t(\w)\equiv -\left[\w-\D\tilde{\e}(\t)+i\tilde{K}^{\t}_{c}(\w)\right]^{-1} 
\circ \D\tilde{f}(\t)
\label{eq:neq_Xo_approx}
\ee
where the four--index tensor $\D\tilde{f}(\t)$ is defined as in 
Eq.~(\ref{dftensor}) with $f_{i}^{\rm eq}\to \tilde{f}_{i}(\t)$.
In addition, the pump-induced renormalization of the single--particle 
energy levels is
\be
\tilde{\e}_i(\t)\approx 
\e^{\rm eq}_i+\blE(\t)\cdot\bld_{ii}+\D\Sigma_{s,ii}^{\t},
\label{eq:bgr}
\ee
where $\D\Sigma_{s}^{\t}$ is a 
time-local functional of the occupations only.
The $\t$-dependent renormalization of the energy levels represents 
the explanation in MBPT language 
of the well known 
band gap renormalization effect, i.e., the reduction of 
the elemental gap induced by pump excited carriers.

Another consequence of the diagonal structure of the density matrix 
is that the static kernel too becomes
a functional of the NEQ occupations only: ${\tilde{K}^{\t}_{s} \approx 
\tilde{K}_{s}[\tilde{f}(\t)]}$. This dependence 
can be used to interpret the renormalization of the electron--hole interaction
and hence, in systems with bound excitons, the renormalization of the excitonic binding energy.

\section{A numerical example}
\label{sec:numerics}

\subsection{Model}
\label{sec:numerics_model}
We illustrate the theory developed in the previous sections by 
calculating the transient photoabsorption spectrum of a 
four-level model system with two valence states 
(orbital quantum numbers $\m=1,2$) and two ``conduction'' or excited states 
(orbital quantum numbers $\m=3,4$). In second quantization the equilibrium 
Hamiltonian reads
\bea
\hat{H}^{\rm eq}= \sum_{\m\s} \e_{\m} \hat{n}_{\m\s}
+ \frac{1}{2} \sum_{\substack{\m\n\\\s\s'}}v_{\m\n}  
\hat{c}^{\dagger}_{\m\s}\hat{c}^{\dagger}_{\n\s'} 
\hat{c}_{\n\s'}\hat{c}_{\m\s},
\eea
with $\hat{n}_{\m\s}=\hat{c}^{\dagger}_{\m\s}\hat{c}_{\m\s}$ the 
occupation operator of level $\m$ with spin $\s$. The system is 
driven out of equilibrium by a strong pulse which pumps electrons 
from the valence states to the conduction states. 
In accordance with the notation of Eqs.~(\ref{eq:H_MB_def}) we 
consider a pump-dipole coupling of the form
\be
\blE(t)\cdot\hat{\bld}=E(t)\sum_{\substack{\m=1,2\\\n=3,4}}\sum_{\s}
\left(d_{\m\n}\hat{c}^{\dagger}_{\m\s}\hat{c}_{\n\s} +{\rm 
H.c.}\right),
\label{PPdcoupl}
\ee
where $d_{\m\n}=\bra \varphi_{\m} | \bgh_{P}\cdot \hat{\blr} 
|\varphi_{\n}\ket  = \bgh_{P} \cdot \bld_{\m\n}$.
After a time $\t$ the excited system is irradiated by a weak 
ultrafast probe. The probe--dipole coupling is the same as in 
Eq.~(\ref{PPdcoupl}) except that the field amplitude $E(t)$ is 
replaced by the amplitude $e(t)$ of the probe pulse. For the 
numerical simulations we choose the 
amplitudes $E(t)$ and $e(t)$ as\cite{blm.2012}
\be
E(t)=E_{0}\sin^{2}\big(\pi\frac{ t}{\D_{P}}\big)\sin \w_{P}t 
\ee
for $0<t<\D_{P}$ and zero otherwise, and 
\be
e(t)=e_{0}\sin^{2}\big(\pi\frac{t-\D_{P}-\tau}{\D_{p}}\big)\sin  
\w_{p}(t-\D_{P}-\tau)
\ee
for $0<t-\D_{P}-\tau<\D_{p}$ and zero otherwise.

The equation of motion for the single-particle density matrix is  
Eq.~(\ref{eq:kbe_general}). For $h^{t}(t)$ we take the HF 
Hamiltonian (see discussion just before Section~\ref{sec:negf_gbka})
\bea
h^{t}_{\m\n}(t)&=&\d_{\m\n}\big[\e_{\m}+\sum_{\a}2v_{\m\a}\r_{\a\a}(t)\big]
-v_{\m\n}\r_{\n\m}(t)
\nn\\&+&[E(t)+e(t)]d_{\m\n}.
\eea
For the collision integral we consider a two-step relaxation 
approximation (in matrix form)
\be
I^{t}(t)\approx \left\{\G^{\rm pol}(t),\r(t)-\r^{\rm qs}\right\}+
\left\{\G^{\rm scatt}(t),\r(t)-\r^{\rm eq}\right\},
\label{modcoll}
\ee
where the curly brackets signify an anticommutator. In 
Eq.~(\ref{modcoll}) the first term accounts for the dephasing 
of the pump-induced polarization and is responsible for driving the 
system toward a
quasi--stationary state described by $\r^{\rm qs}$. After the dephasing 
$\G^{\rm pol}(t)\approx 0$ and the collision integral is dominated by the second 
term which describes the relaxation toward the equilibrium state. 
The damping matrices $\G^{\rm pol}(t)=\g^{\rm pol}(t)\1$ and $\G^{\rm 
scatt}(t)=\g^{\rm scatt}(t)\1$ are proportional to the identity 
matrix, thus guaranteeing the conservation of the total number of 
particles $N=2\Tr[\r]$. Since there is no pump-induced dephasing in the absence 
of the pump  $\G^{\rm pol}$ is proportional to the amplitude of the pump pulse.

\begin{figure}[tbp]
\includegraphics[width=8.5cm]{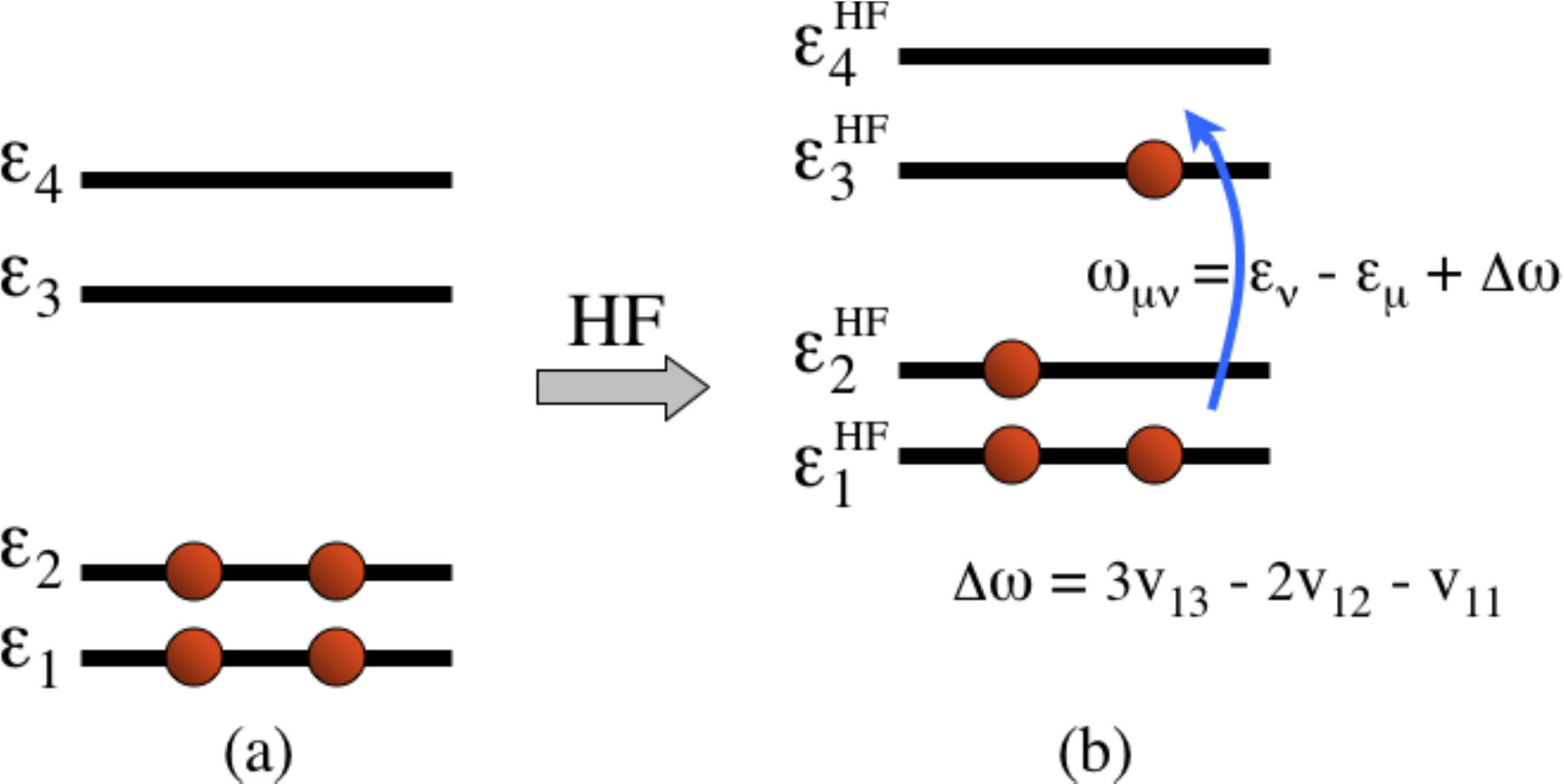}
\includegraphics[width=8.5cm]{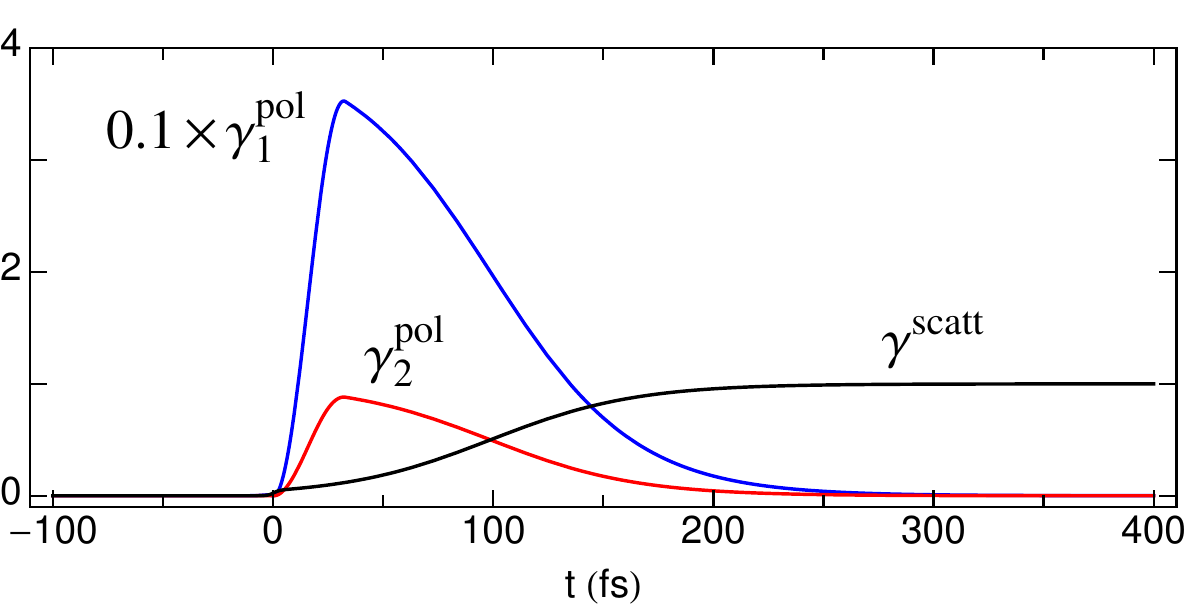}
\caption{Upper panel: (a)  
non--interacting energy levels $\e_{1}=0$, $\e_{2}=0.1$, $\e_{3}=1.0$, 
$\e_{4}=1.3$ and (b) HF energy levels $\e_{\m}^{\rm HF}=\e_{\m}
+\sum_{\a}f^{\rm eq}_{\a}(2v_{\m\a}-\d_{\m\a}v_{\m\m})$ with  
$v_{\m\m}=0.4$, $v_{12}=v_{23}=0.2$ and otherwise $v_{\mu\nu}=0.1$ 
(we recall that $f^{\rm eq}_{\a}=\r^{\rm eq}_{\a\a}$). 
The position of the poles $\w_{\m\n}$ of the equilibrium TD-HF 
$\chi^{\rm eq}(\w)$ 
(solution of Eqs.~(\ref{chi0eq}-\ref{chieq}) with $K_{c}^{\rm eq}=0$) can be 
calculated analytically and are 
also indicated.
Lower panel: Plot of the damping functions $\g^{\rm scatt}(t)$, 
$\g^{\rm pol}_{1}(t)$ (reduced by a 
factor of 10) and $\g^{\rm pol}_{2}(t)$ used in the numerical 
simulations. Energies $\e_{\m},\e^{\rm HF}_{\m}$ and $v_{\m \n}$ are in 
eV while  $\g^{\rm scatt}$ and $ \g^{\rm pol}_{1,2}$ are in meV.}
\label{levelscheme}
\end{figure}

The system has  
filled valence states and empty conduction states at time $t=0$, 
hence $\r^{\rm eq}={\rm diag}\{ 
1,1,0,0\}$. The 
model parameters as well as the HF equilibrium configuration can be 
found in Fig.~\ref{levelscheme}. For the dipole matrix we use
\be
d = d_{0}\left(
\begin{array}{cccc}
0 & 0 & 1 & 1 \\
0 & 0 & 1 & 1 \\
1 & 1 & 0 & 0 \\
1 & 1 & 0 & 0 
\end{array}
\right).
\ee
The damping functions $\g^{\rm scatt}$ and the 
two different $\g^{\rm pol}=\g^{\rm pol}_{1},\g^{\rm pol}_{2}$ that we consider
are illustrated in the lower panel of Fig. \ref{levelscheme}. 
In particular $\g^{\rm pol}$ is responsible for the relaxation toward the 
quasi--stationary density matrix $\r^{\rm qs}={\rm diag}\{ 
0.9,0.9,0.1,0.1\}$.
For the external fields we study a pump pulse of 
duration $\D_{P}=66$ fs
and frequency 
$\w_{P}=0.6$ eV, and a probe of duration $\D_{p}=20$ fs
and frequency 
$\w_{p}=0.6$ eV; the amplitudes  $E_{0}$, $e_{0}$ and $d_{0}$ are 
chosen to yield $E_{0}d_{0}=0.1$ eV and $e_{0}d_{0}=0.001$ eV.

We calculate $\r(t)$ in the presence of both pump and probe as well as the 
probe--free $\tilde{\r}$ and then extract the probe--induced dipole moment 
$\d d(t)=d_{\m\n}(\r_{\n\m}(t)-\tilde{\r}_{\n\m}(t))$.
Successively, we obtain the 
transient spectrum of Eq.~(\ref{eq:trans_spectrum}) by Fourier transforming the function $\d 
d(t) \times e^{-t/\tilde{\tau}_{p}}$ with $\tilde{\tau}_{p}= 80$ fs the 
life--time of the probe--induced dipole. In the figures below the 
exponential damping is always included in the probe--induced dipole.  
The probe--free 
$\tilde{\r}$ is also used in Eq.~(\ref{eq:kbe_lin_chi_adiab}) to 
calculate the adiabatic NEQ response function and hence the transient 
spectrum according to Eq.~(\ref{transspectrum}). The quality of the 
adiabatic approximation is assessed in different regimes.

\begin{figure}[tbp]
\includegraphics[width=8.5cm]{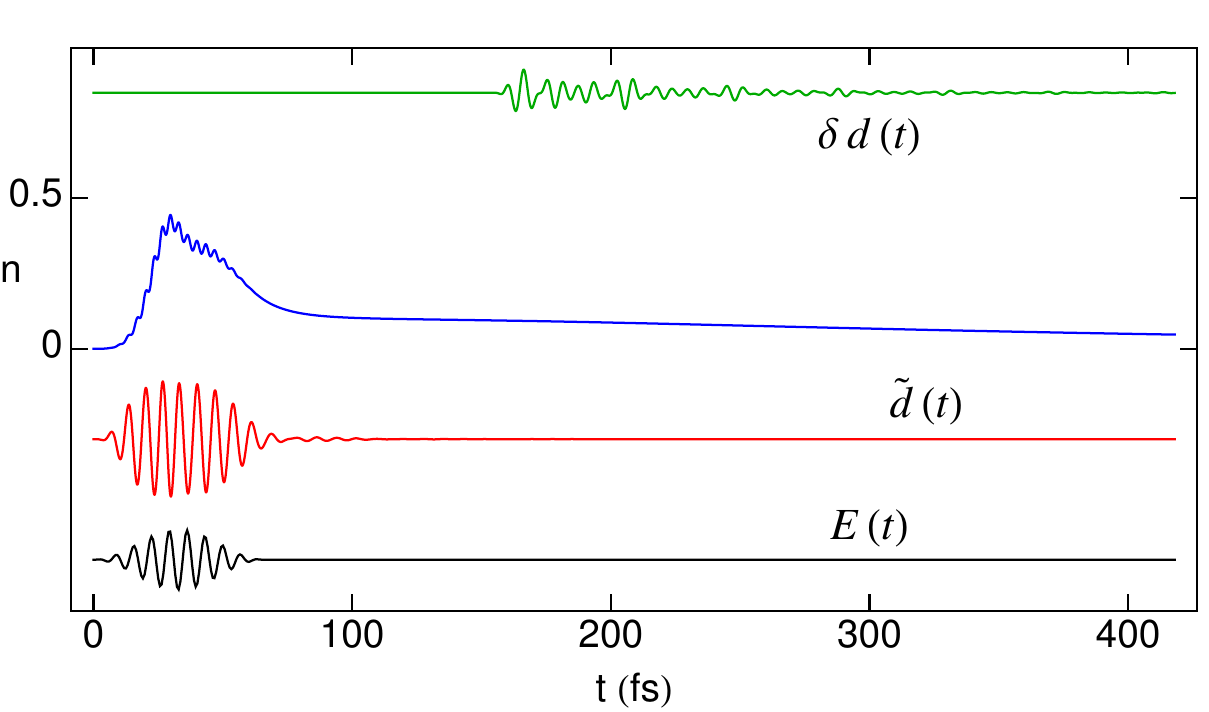}\\
\includegraphics[width=8.5cm]{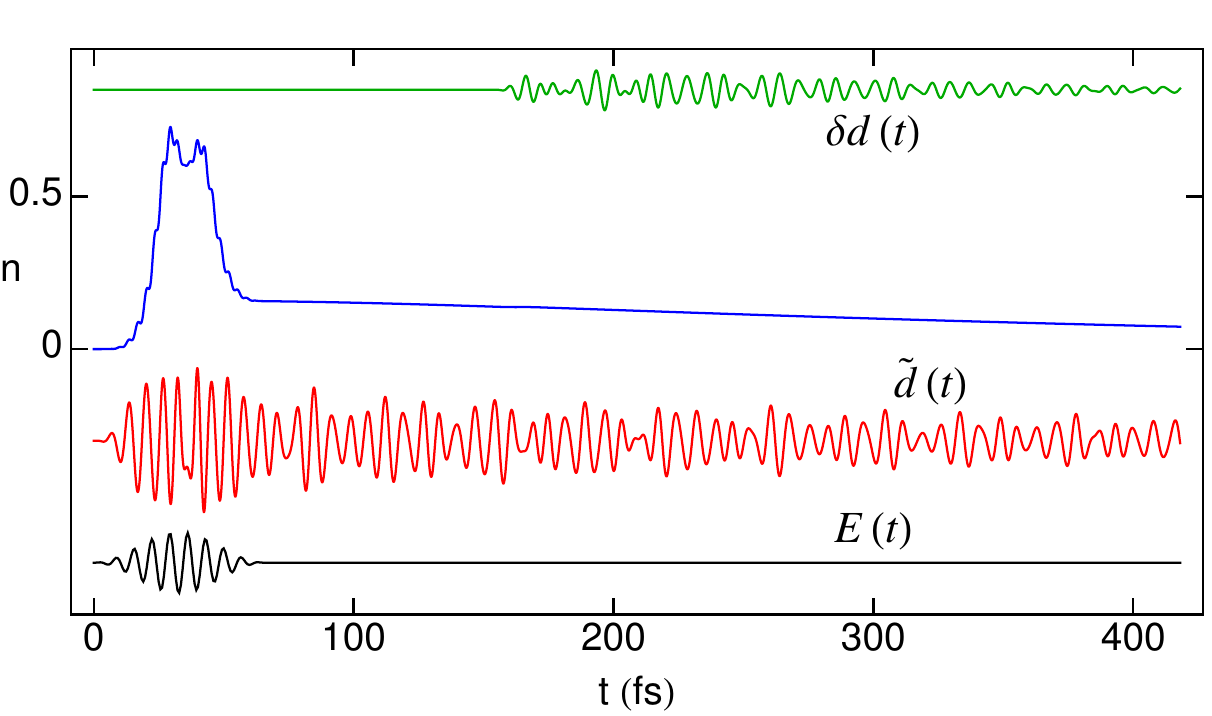}
\caption{From bottom to top, pump pulse $E(t)$ (black), probe--free dipole 
$\tilde{d}(t)$ (red), occupation $n_{3}=n_{4}\equiv n$ of valence states 
(blue) and probe--induced dipole $\d d(t) $ (green) for the large 
$ \g^{\rm pol}_{1}$ (top panel) and small $ \g^{\rm pol}_{2}$ (bottom 
panel) damping functions. These results 
are obtained for a delay $\tau=130$ fs. The quantities $E(t)$, 
$\tilde{d}(t)$ and $\d d(t) $ are in arbitrary units. }
\label{fig3}
\end{figure}
\subsection{Results and discussion}
In the upper panel of Fig. \ref{fig3} 
we show some relevant quantities obtained from the numerical solution of 
Eq. (\ref{eq:kbe_general}) with $\g^{\rm pol}=\g^{\rm pol}_{1}$, namely (from 
bottom to top) the pump pulse $E(t)$, the probe--free dipole 
$\tilde{d}(t)$, the time-dependent
occupation $n_{3}(t)=n_{4}(t)\equiv n(t)$ of the valence states 3 and 4,  and the
probe--induced dipole $\d d(t)$.
The behavior of these quantities resemble the behavior in Fig. 
\ref{timescales}. When the probe arrives ($\tau = 130$ fs), the pump-induced polarization is 
completely dephased  ($\t_{\rm pol}\sim 100$ fs), and the system is slowly moving
around the  quasi--stationary excited state described by 
$\r^{\rm qs}$. In this situation the time-scale over which the 
one--particle density matrix changes is  $T_{P}\sim 1/\g^{\rm 
scatt}\sim 10^{3}$ fs. Since $T_{P}$ is much larger than the life--time 
$\t_{p}=\tilde{\t}_{p}+\D_{p}= 100$ fs of the dressed probe the adiabatic 
condition is fulfilled, see Eq.~(\ref{eq:ad_ansatz}).

\begin{figure}[tbp]
\includegraphics[width=8.5cm]{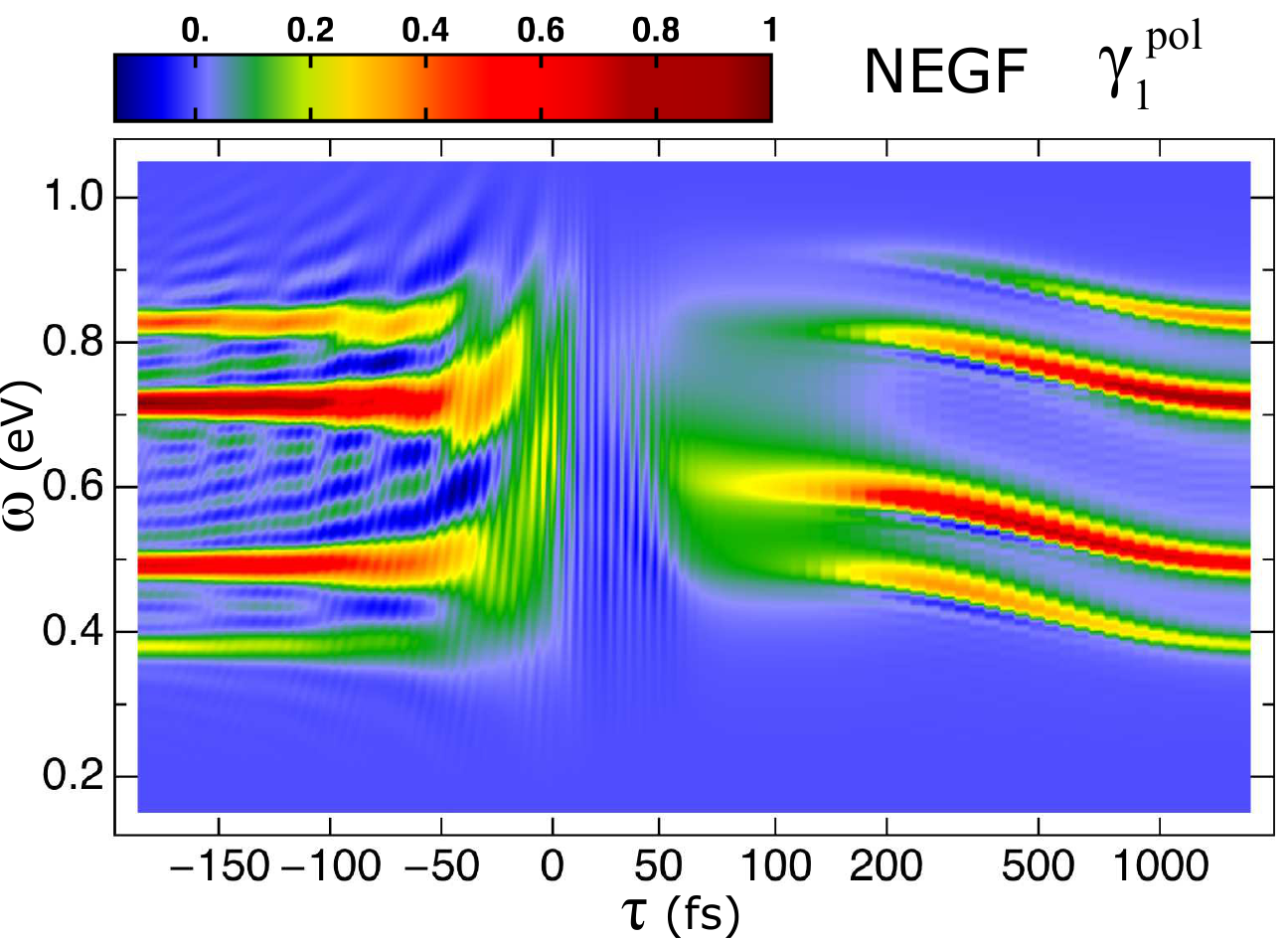}\\
\includegraphics[width=8.5cm]{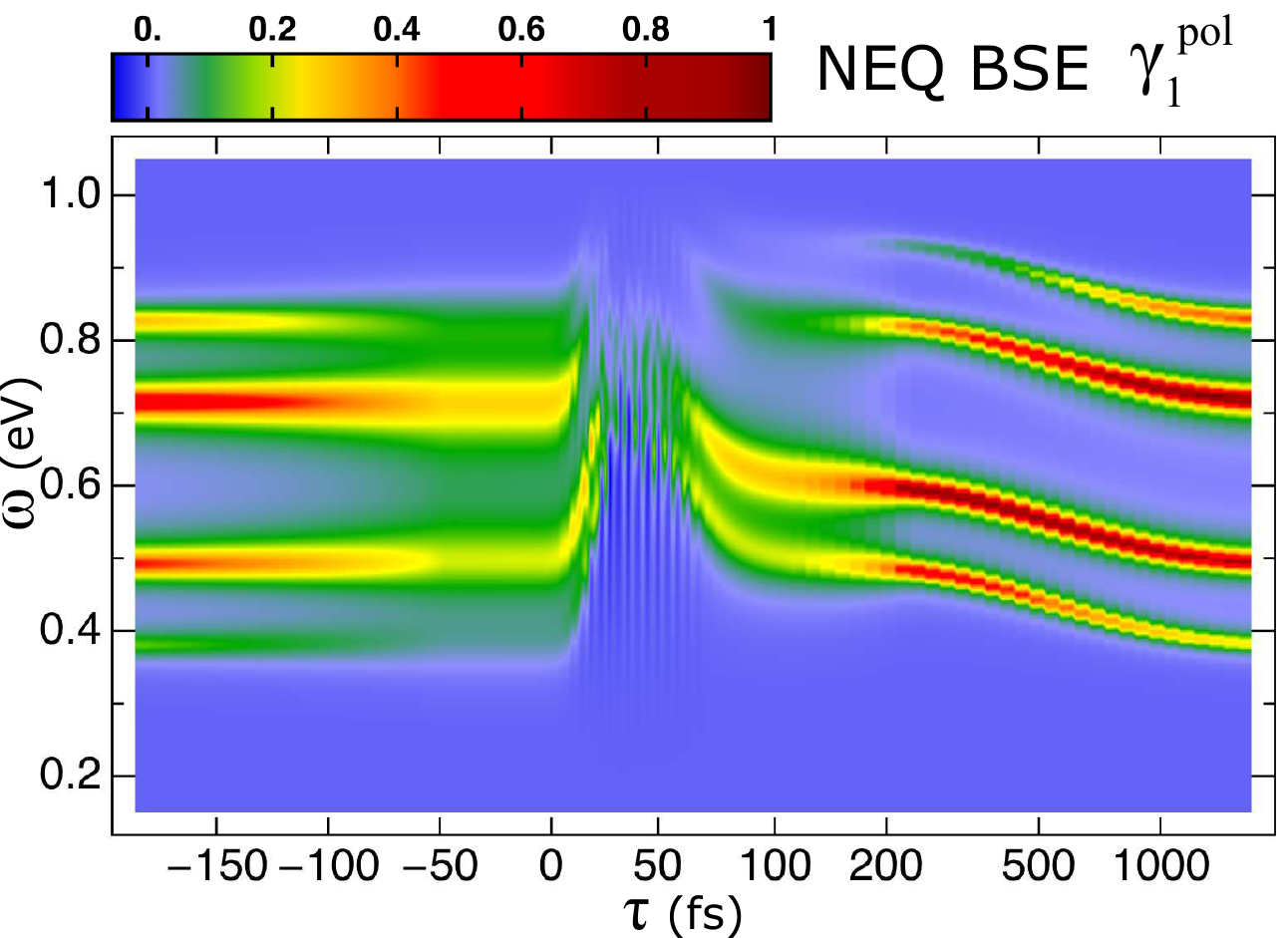}
\caption{Transient absorption  spectrum 
$\mathfrak{S}^{\t}(\w)$ (normalized to its maximum value)
obtained within NEGF according to Eq. (\ref{eq:trans_spectrum}) (upper panel) and
with the NEQ--BSE  according to  Eq. (\ref{transspectrum}) (lower 
panel) using the damping function $\g^{\rm pol}=\g^{\rm pol}_{1}$. 
The peaks of the NEQ--BSE spectrum have been 
broadened by the inverse life--time $1/\tilde{\t}_{p}$ of $\delta d(t)$.
Notice that the scale of the horizontal axis is linear for $\tau<100$ fs and 
logarithmic otherwise. }
\label{fig4}
\end{figure}

The transient absorption spectra $\mathfrak{S}^{\t}(\w)$
obtained within NEGF according to Eq. (\ref{eq:trans_spectrum})  and
with the NEQ--BSE according to  Eq. (\ref{transspectrum}) 
are displayed in Fig. \ref{fig4}.
As expected, the NEQ--BSE approach is very accurate for delays 
$\t \lesssim -\t_{p}$ and  $\t \gtrsim \t_{\rm pol}$, i.e., when 
the probe--induced dipole does not overlap the pump-induced 
polarization. For $\t \lesssim -\t_{p}$ the spectrum 
$\mathfrak{S}^{\t}(\w)$ is the 
equilibrium spectrum  with four peaks at energies 
$\w_{\m\n}=\e_{\n}-\e_{\m}+\D\w$ (with 
$\m=1,2$, $\n=3,4$ and $\D\w = 3v_{13}-2v_{12}-v_{11}$), thus  
NEGF and NEQ--BSE obviously agree.
For $\t \gtrsim \t_{\rm pol}$ the system is in a nonequilibrium state
and the condition of adiabaticity matters. 
The NEQ--BSE well captures the $\tau$-dependent 
structure of the NEGF spectrum, with the correct bending of 
the position of the four main absorption peaks towards their 
equilibrium value for large $\t$.
At first sight, the agreement seems rather good in the overlap region 
$-\t_{p}<\tau< \t_{\rm pol}$ too.
However, in this region $\mathfrak{S}^{\t}(\w)$ is very small
due to the sizable broadening induced by the large $\g^{\rm 
pol}_{1}$, and a more careful comparison between the NEQ--BSE and NEGF 
spectra reveals some discrepancies (not shown). 

\begin{figure}[tbp]
\includegraphics[width=8.5cm]{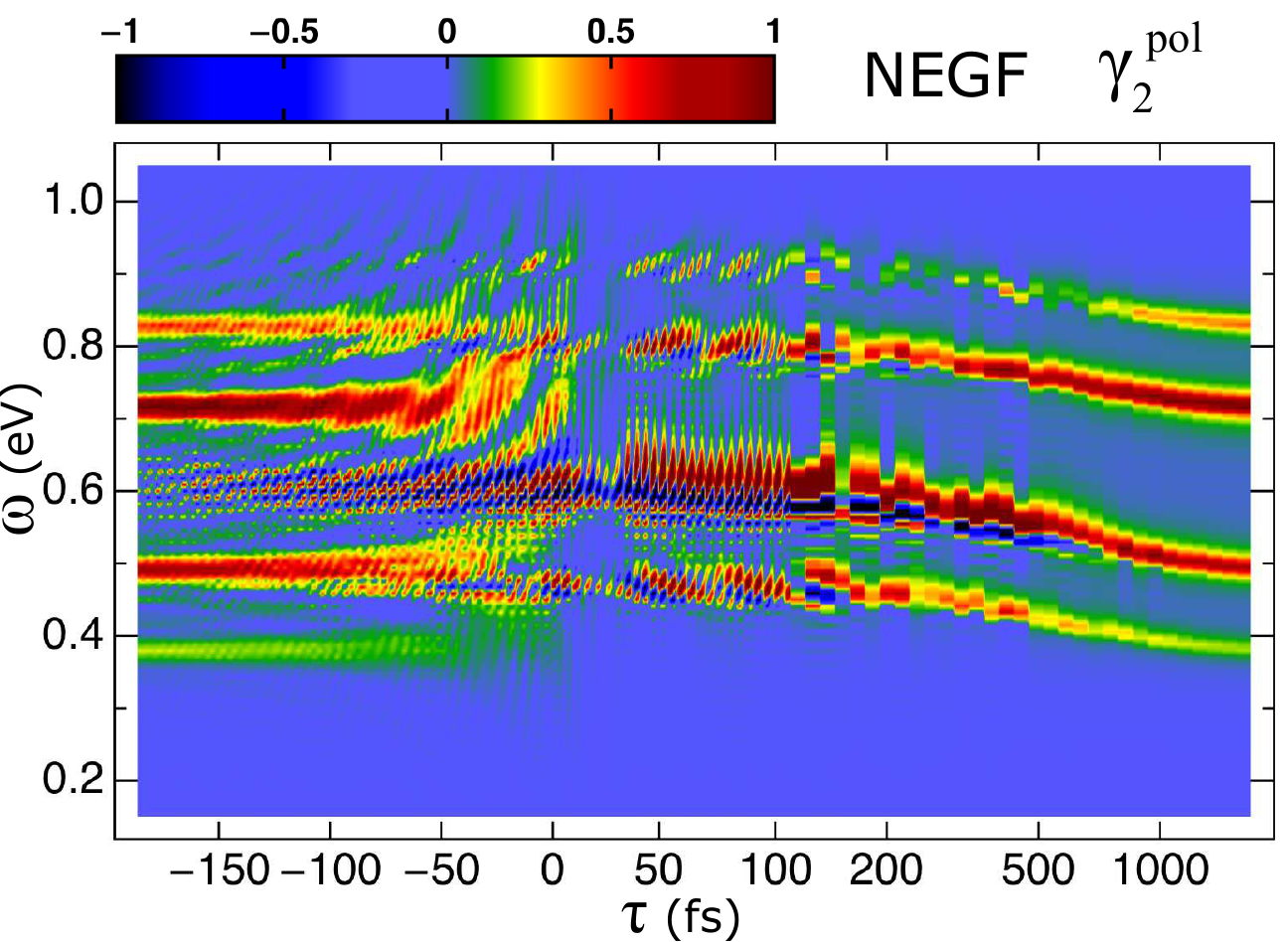}\\
\includegraphics[width=8.5cm]{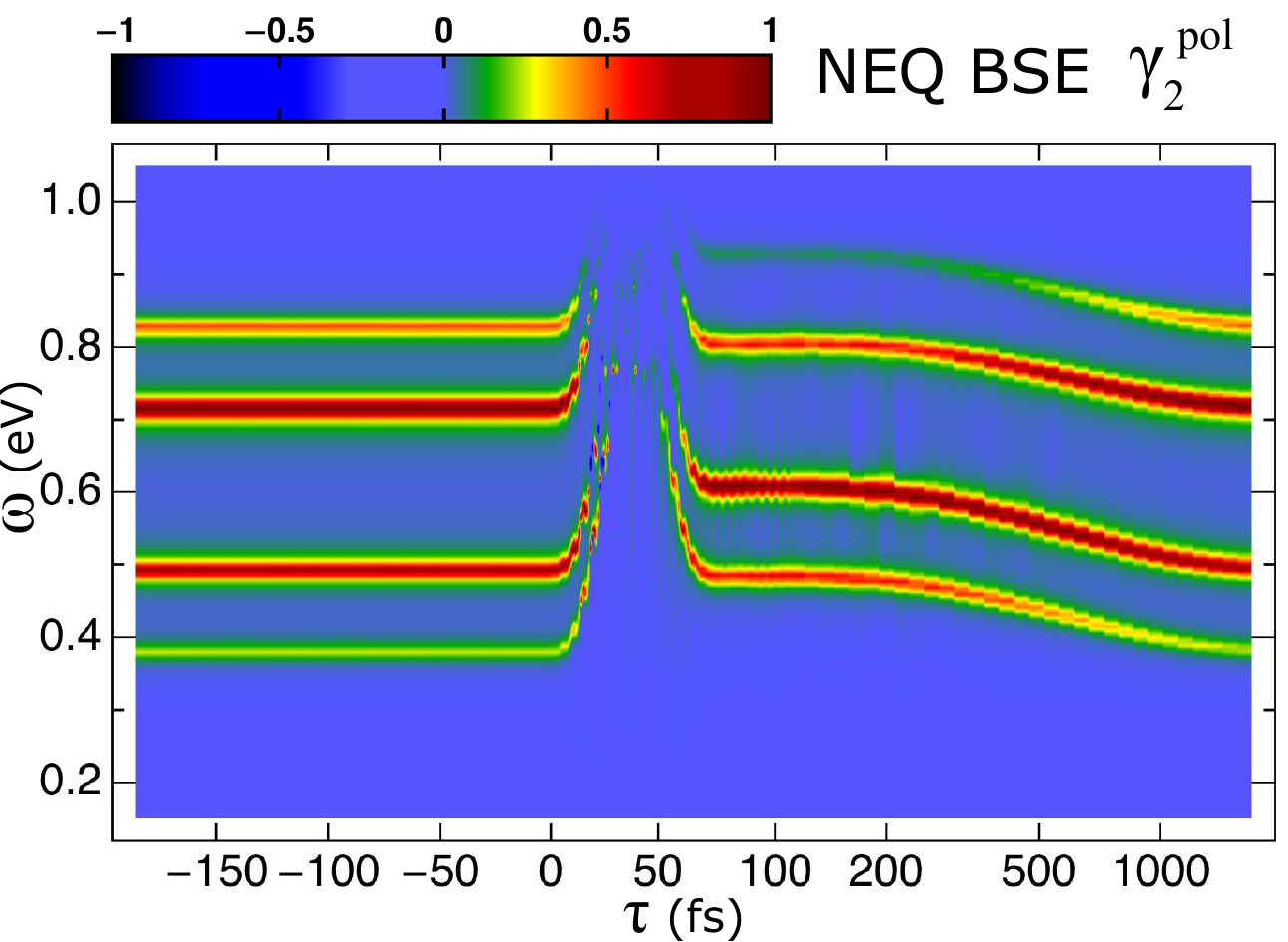}
\caption{Same as Fig. \ref{fig4} except that the small damping function $ \g^{\rm 
pol}_{2}$ has been used in Eq. (\ref{modcoll}). }
\label{fig5}
\end{figure}

A second simulation has been performed using the smaller damping 
function $\g^{\rm pol}=\g^{\rm pol}_{2}$. In this case the pump-induced polarization $\tilde{d}(t)$ is 
long--lived, as it can be seen in the lower panel of Fig. \ref{fig3}. 
After a time $\t=130$ fs the time--scale $T_{P}$ over which 
the one--particle density matrix changes is given by the period of the coherent 
oscillations of $\tilde{d}(t)$ and it is roughly equal 
to the inverse gap $1/\w_{23}\approx 10$ fs.
Thus the condition of adiabaticity $T_{P} \gg \t_{p}$ is not 
fulfilled and no agreement between  NEQ--BSE and NEGF is expected.
The transient absorption spectra are 
displayed in Fig.~\ref{fig5} showing that the two approaches 
differ whenever the probe experiences a sizable  
$\tilde{d}(t)$, i.e., for $-\t_{p} < \t \lesssim 500$.
In this region
the NEGF spectrum exhibits alternating fringes characterized by a large 
oscillation of the spectral weight at fixed $\w$ as a function of $\t$.
These features origin from the nonadiabatic coherent motion of the
electrons between valence and conduction states,
and hence they are out of reach of the NEQ--BSE approach. 
Remarkably, however, the NEQ--BSE captures important spectral 
features even in this strongly nonadiabatic situation, the most 
prominent feature being the upward bending of the main peaks around 
$\t=0$.  When the coherence is destroyed by the dephasing, i.e., for 
$\t>500$ fs, the NEQ--BSE and NEGF spectra are found to be in 
excellent agreement.

\section{Conclusions}
\label{sec:conclusions}

We propose a practical method based on MBPT to calculate  
\ppe spectra for delays in the ``adiabatic'' regime.
Starting from the KBE for the Keldysh Green's function we use the 
GKBA to obtain an equation of motion for the one--particle density 
matrix $\r$ in the presence of both pump and probe fields. 
Linearization around zero probe yields an equation 
for the NEQ response function $\chi(t,t')$. After the action of the 
pump we identify a physically 
relevant regime during which the probe--free 
density matrix $\tilde{\r}$ varies on a time-scale much longer than 
the life--time of the dressed probe. In this regime we make the 
adiabatic approximation and show that $\chi(t,t')$ can be written as 
a function of the pump-probe delay $\t$ and of the relative time
$(t-t')$, i.e., $\chi(t,t')\approx \chi^{\t}(t-t')$. This 
simplification allows 
us to Fourier transform with respect to the relative time and to derive 
the main result of this work: a NEQ--BSE that can be implemented in most
of the \ai\, numerical schemes and codes. We further provide a sound 
physical interpretation of the NEQ response function and showed that 
it can be related to {\em intrinsic} spectral properties of the {\em 
nonequilibrium} system. Well known effects like the renormalization 
of the band--gap and excitonic binding energies in semiconductors 
and insulators are naturally explained.

The computational advantage of the NEQ--BSE over NEGF simulations is 
enormous as  only the probe--free 
one--particle density matrix $\tilde{\r}$ enters in the solution of 
the NEQ--BSE. This implies that a single time--propagation
is sufficient to obtain the transient spectrum for several delays. In 
contrast, the NEGF approach requires a time-propagation for every 
delay (to obtain the one--particle density matrix with pump and probe fields)
in addition to the time--propagation to obtain $\tilde{\r}$. The 
validity of the NEQ--BSE has been successfully demonstrated in a simple four-level 
model system and it is currently under investigation in more realistic 
hamiltonians with encouraging results.

\section*{Acknowledgments}
We acknowledge financial support by the {\em Futuro in Ricerca} grant No. RBFR12SW0J of the
Italian Ministry of Education, University and Research MIUR. AM and DS also acknowledge
the funding received from the European Union Horizon 2020 research and
innovation program under grant agreement No 654360.


\begin{thebibliography}{10}
    
\bibitem{bgk.2009}
R. Berera, R. van Grondelle and J. T. M. Kennis,
Photosynth. Res. {\bf 101}, 105 (2009). 
    
\bibitem{spsk.2012}
G. Sansone, T. Pfeifer, K. Simeonidis and A. I. Kuleff,
Chem. Phys. Chem. {\bf 13}, 661 (2012). 

\bibitem{ghlsllk.2013}
L. Gallmann, J. Herrmann, R. Locher, M. Sabbar, A. Ludwig, M. 
Lucchini and U. Keller,
Mol. Phys. {\bf 111}, 2243 (2013).

\bibitem{kc.2014}
A. I. Kuleff and L. S. Cederbaum,
J. Phys. B: At. Mol. Opt. Phys. {\bf 47}, 124002 (2014).  

\bibitem{science.2014}
S. M. Falke, C. A. Rozzi, D. Brida, M. Maiuri,
M. Amato, E. Sommer, A. De Sio, A. Rubio, G. Cerullo, E. Molinari and 
C. Lienau,
Science {\bf 344}, 1001 (2014).

\bibitem{bm.1997}
W. Beenken and V. May,
J. Opt. Soc. Am. B {\bf 14}, 2804 (1997).

\bibitem{wssd}
B. Wolfseder, L. Seidner, G. Stock and W. Domcke,
Chem. Phys. {\bf 217}, 275 (1997).

\bibitem{sypl.2011}
R. Santra, V. S. Yakovlev, T. Pfeifer and Z. Loh,
Phys. Rev. A {\bf 83}, 033405 (2011).

\bibitem{mpb.2012}
A. S. Moskalenko, Y. Pavlyukh and J. Berakdar,
Phys. Rev. A {\bf 86}, 013202 (2012).

\bibitem{blm.2012}
J. C. Baggesen, E. Lindroth and L. B. Madsen,
Phys. Rev. A {\bf 85}, 013415 (2012).

\bibitem{dgc.2013}
A. D. Dutoi, K. Gokhberg and L. S. Cederbaum,
Phys. Rev. A {\bf 88}, 013419 (2013).

\bibitem{PS.2015}
E. Perfetto and G. Stefanucci,
Phys. Rev. A {\bf 91}, 033416 (2015).

\bibitem{Shen-book}
Y. Shen, {\em The principles of nonlinear optics} (Wiley Series in 
Pure and Applied Optics, J. Wiley, 1984).

\bibitem{kbs.1988}
G. Khitrova, P. R. Berman and M. Sargent III,
J. Opt. Soc. Am. B {\bf 5}, 160 (1988).

\bibitem{ym.1990}
Y. J. Yan and S. Mukamel,
Phys. Rev. A {\bf 41}, 6485 (1990).

\bibitem{pm.1992}
W. T. Pollard and R. A. Mathies,
Annu. Rev. Phys. Chem. {\bf 43}, 497 (1992).

\bibitem{Mukamel-book}
S. Mukamel, {\em Principles of Nonlinear Optical Spectroscopy} 
(Oxford University Press, Oxford, 1995).

\bibitem{yzc.1997}
Y. J. Yan, W. Zhang and J. Che,
J. Che. Phys. {\bf 106}, 2212 (1997).

\bibitem{strinati_review}
G. Strinati,
Rivista del Nuovo Cimento {\bf 11}, 1 (1988).

\bibitem{svl-book}
G. Stefanucci and R. van Leeuwen,
{\em Nonequilibrium Many-Body Theory of Quantum Systems: A Modern Introduction}
(Cambridge University Press, Cambridge, 2013).

\bibitem{Strinati}
G. Strinati, H. J. Mattausch and W. Hanke,
Phys. Rev. B {\bf 25}, 2867 (1982).

\bibitem{orr.2002}
G. Onida, L. Reining, and A. Rubio,
Rev. Mod. Phys. {\bf 74}, 601 (2002).

\bibitem{ardo.1998} 
S. Albrecht, L. Reining, R. Del Sole, G. Onida, 
Phys. Rev. Lett. {\bf 80}, 4510 (1998).

\bibitem{bsb.1998}
L. X. Benedict, E. L. Shirley, R. B. Bohn, 
Phys. Rev. Lett. {\bf 80}, 4514 (1998).

\bibitem{rl.1998}
M. Rohlfing, S. G. Louie, 
Phys. Rev. Lett. {\bf 81}, 2312 (1998)

\bibitem{pphs.2011}
G. Pal, Y. Pavlyukh, W. H\"ubner, and H.C. Schneider,
Eur. Phys. J. B {\bf 79}, 327 (2011).

\bibitem{pgg.1996}
M. Petersilka, U. J. Gossmann and E. K. U. Gross,
Phys. Rev. Lett. {\bf 76}, 1212 (1996).

\bibitem{Casida1}
M. E. Casida,
in Recent Advances in Density Functional Methods, Part I, edited by 
D.P. Chong (Singapore, World Scientific, 1995), p. 155.

\bibitem{rg.1984}
E. Runge and E. K. U. Gross,
Phys. Rev. Lett. {\bf 52}, 997 (1984).

\bibitem{Ullrich}
C. A. Ullrich, {\em Time-Dependent Density-Functional Theory: 
Concepts and Applications} (Oxford University Press, 2011).

\bibitem{rozzi-nature}
C. A. Rozzi, S. M. Falke, N. Spallanzani, A. Rubio, E. Molinari, D. 
Brida, M. Maiuri, G. Cerullo, H. Schramm, J. Christoffers and C. Lienau, 
Nat. Comm. {\bf 4}, 1602 (2013).

\bibitem{miguel}
Ch. Neidel, J. Klei, C.-H. Yang, A. Rouz\'ee, M. J. J. Vrakking, K. 
Kl\"under, M. Miranda, C. L. Arnold, T. Fordell, A. L'Huillier, M. 
Gisselbrecht, P. Johnsson, M. P. Dinh, E. Suraud, P.-G. Reinhard, V. 
Despr\'e, M. A. L. Marques, and F. L\'epine,
Phys. Rev. Lett. {\bf 111}, 033001 (2013).

\bibitem{degiovannini}
U. De Giovannini, G. Brunetto, A. Castro, J. Walkenhorst and A. Rubio,
ChemPhysChem {\bf 14}, 1363 (2013).

\bibitem{mzcb.2004}
N. T. Maitra, F. Zhang, R. J. Cave and K. Burke,
J. Chem. Phys. {\bf 120}, 5932 (2004).

\bibitem{kk.2008}
S. K\"ummel and L. Kronik,
Rev. Mod. Phys. {\bf 80}, 3 (2008). 

\bibitem{ngb.2006}
O. Gritsenko and E. J. Baerends,
J. Chem. Phys. {\bf 121}, 655 (2004).

\bibitem{m.2005}
N. T. Maitra,
J. Chem. Phys. {\bf 122}, 234104 (2005).

\bibitem{mt.2006}
N. T. Maitra and D. G. Tempel,
J. Chem. Phys. {\bf 125}, 184111 (2006).

\bibitem{sk.2011}
G. Stefanucci and S. Kurth,
Phys. Rev. Lett. {\bf 107}, 216401 (2011).

\bibitem{ks.2013}
S. Kurth and G. Stefanucci,
Phys. Rev. Lett. {\bf 111}, 030601 (2013).

\bibitem{roro.2002}
L. Reining, V. Olevano, A. Rubio and G. Onida,
Phys. Rev. Lett. {\bf 88}, 066404 (2002).

\bibitem{nature.2010}
E. Goulielmakis, Z. Loh, A. Wirth, R. Santra, N. Rohringer,
V. S. Yakovlev, S. Zherebtsov, T. Pfeifer, A. M. Azzeer, M. F. Kling, 
S. R. Leone and F. Krausz, 
Nature {\bf 466}, 739 (2010).

\bibitem{rs.2009}
N. Rohringer and R. Santra,
Phys. Rev. A {\bf 79}, 053402 (2009).

\bibitem{hj-book}
H. Haug and A.-P. Jauho,
{\em Quantum Kinetics in Transport and Optics of Semiconductor} 
(Springer-Verlag, Berlin, 1998).

\bibitem{footnote_thin} Here for optically thin we mean that the sample is microscopic
                        along the propagation direction of the light-pulse.

\bibitem{vLDSAvB.2006}
R. van Leeuwen, N. E. Dahlen, G. Stefanucci, C.-O. Almbladh and U. von Barth,
Lect. Notes Phys. {\bf 706}, 33 (2006).

\bibitem{d.1984}
P. Danielewicz,
Ann. Phys. {\bf 152}, 239 (1984).

\bibitem{bb.2013}
K. Balzer and M. Bonitz,
{\em Nonequilibrium Green's Functions Approach to Inhomogeneous Systems}, 
Lecture Notes in Physics
(Springer-Verlag, Berlin Heidelberg, 2013), Vol. 867.

\bibitem{kb-book}
L. P. Kadanoff and G. Baym,
{\em Quantum Statistical Mechanics}
(Westview Press, Boulder, CO, 1994).

\bibitem{lsv.1986}
P. Lipavsk\'y, V. $\check{\rm S}$pi$\check{\rm c}$ka and B. Velick\'y,
Phys. Rev. B {\bf 34}, 6933 (1986).

\bibitem{FLSM.2015}
J. I. Fuks, K. Luo, E. D. Sandoval, and N. T. Maitra,
Phys. Rev. Lett. {\bf 114}, 183002 (2015).

\bibitem{DvL.2006}
N. E. Dahlen and R. van Leeuwen,
Phys. Rev. Lett. {\bf 98}, 153004 (2007).

\bibitem{MSSvL.2009}
P. My\"oh\"anen, A. Stan, G. Stefanucci and R. van Leeuwen,
Phys. Rev. B {\bf 80}, 115107 (2009).




\bibitem{marini.2008} 
A. Marini,
Phys. Rev. Lett. {\bf 101}, 106405 (2008).

\bibitem{marini.2003} 
A. Marini and R. Del Sole,
Phys. Rev. Lett. {\bf 91}, 176402 (2003).








\bibitem{h.1992}
H. Haug,
Phys. Status Solidi B {\bf 173}, 139 (1992).

\bibitem{bsh.1999}
M. Bonitz, D. Semkat, and H. Haug,
Eur. Phys. J. B {\bf 9}, 309 (1999).

\bibitem{m.2012}
A. Marini, 
J. Phys. Conf. Ser. {\bf 427}, 012003 (2013).

\bibitem{LPUvLS.2014}
S. Latini, E. Perfetto, A.-M. Uimonen, R. van Leeuwen and G. 
Stefanucci,
Phys. Rev. B {\bf 89}, 075306 (2014). 

\bibitem{Att.2011}
C. Attaccalite, M. Gr\"{u}ning, A. Marini,
Phys. Rev. B {\bf 84}, 245110 (2011). 

\end{thebibliography}
\end{document}